\documentclass[journal]{IEEEtran}

\usepackage{color}

\usepackage{xspace}
\usepackage{graphicx}
\usepackage{textcomp}
\usepackage[para]{footmisc}
\usepackage{xcolor}
\usepackage{hanging}
\usepackage{subfig}

\usepackage{amsmath}
\usepackage{caption}

\usepackage{float}
\usepackage{subfiles}

\usepackage{hyperref}

\def\hyphenateAndTtWholeString #1{\xHyphenate#1$\wholeString\unskip}

\def\xHyphenate#1#2\wholeString {\if#1$%
    \else\transform{#1}%
    \takeTheRest#2\ofTheString\fi}

\def\takeTheRest#1\ofTheString\fi
{\fi \xHyphenate#1\wholeString}

\def\transform#1{\url{#1}\hskip 0pt plus 0pt}

\def\urlx #1{\href{#1}{\hyphenateAndTtWholeString{#1}}}

\newcommand{\project}[0] {{\sffamily\textsc{Bike2Work}}\xspace}
\usepackage{color}      
\usepackage{amssymb}    

\definecolor{ao(english)}{rgb}{0.9, 0.0, 0.1}

\newtheorem{definition}{Definition}

\begin{document}

\title{Play\&Go Corporate: An End-to-End Solution for Facilitating Urban Cyclability}
\color{black}

\author{
Antonio~Bucchiarone \IEEEauthorrefmark{1}\href{https://orcid.org/0000-0003-1154-1382}{\includegraphics[scale=.06]{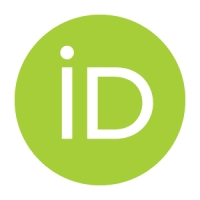}},
Simone Bassanelli\IEEEauthorrefmark{1}\href{https://orcid.org/0000-0001-6061-8169}{\includegraphics[scale=.06]{img/orcid.png}}, Massimiliano Luca\IEEEauthorrefmark{1}\href{https://orcid.org/0000-0001-6964-9877}{\includegraphics[scale=.06]{img/orcid.png}}, Simone Centellegher\IEEEauthorrefmark{1}\\Piergiorgio Cipriano\IEEEauthorrefmark{2},
Luca Giovannini\IEEEauthorrefmark{2}, Bruno Lepri\IEEEauthorrefmark{1}, Annapaola Marconi\IEEEauthorrefmark{1}\href{https://orcid.org/0000-0001-8699-7777}{\includegraphics[scale=.06]{img/orcid.png}}

\vspace{0.5em}
\IEEEauthorrefmark{1}Fondazione Bruno Kessler, Trento, Italy\\
\{bucchiarone,sbassanelli,mluca,centellegher,lepri,marconi\}@fbk.eu\\
\IEEEauthorrefmark{2}DedaNext, Trento, Italy\\
{\{piergiorgio.cipriano,luca.giovannini\}@dedagroup.it}\\
}

%



\maketitle

\begin{abstract}
Mobility plays a fundamental role in modern cities. How citizens experience the urban environment, access city core services, and participate in city life, strongly depends on its mobility organization and efficiency. The challenges that municipalities face are very ambitious: on the one hand, administrators must guarantee their citizens the right to mobility and to easily access local services; on the other hand, they need to minimize the economic, social, and environmental costs of the mobility system. 
Municipalities are increasingly facing problems of traffic congestion, road safety, energy dependency and air pollution, and therefore encouraging a shift towards sustainable mobility habits based on active mobility is of central importance.  Active modes, such as cycling, should be particularly encouraged, especially for local recurrent journeys (e.g., home--to--school, home--to--work). 
In this context, addressing and mitigating commuter-generated traffic requires engaging public and private stakeholders through innovative and collaborative approaches that focus not only on supply (e.g., roads and vehicles) but also on transportation demand management. 
In this paper, we present an end-to-end solution, called  Play\&Go Corporate, for enabling urban cyclability and its concrete exploitation in the realization of a home-to-work sustainable mobility campaign (i.e., \project) targeting employees of public and private companies.
To evaluate the effectiveness of the proposed solution we developed two analyses: the first to carefully analyze the user experience and any behaviour change related to the \project mobility campaign, and the second to demonstrate how exploiting the collected data we can potentially inform and guide the involved municipality (i.e., Ferrara, a city in Northern Italy) in improving urban cyclability.

\end{abstract}


\IEEEpeerreviewmaketitle

\section{Introduction}
\label{sec:intro}

The traditional and car-centric transportation planning has not only significantly contributed to increasing greenhouse emissions but has also detrimentally influenced air quality, traffic congestion, fatalities, and road injuries \cite{su12198170,YANG2020102252}. With the world facing a climate crisis, there is a need for a paradigmatic shift towards a more sustainable and active mobility \cite{Dogan2020}.
Sustaining and attaining significantly higher rates in cycling and walking as modes of transportation (also known as \textit{active mobility}), and nudging people to walk and cycle more, represent one of the most powerful and significant instruments to achieve social goals and sustainability \cite{GARGIULO2022552}.

\textit{Urban cyclability} is a wide notion that connects several elements to bicycle riding, depending on the research's emphasis and interest \cite{Nielsen2018}. It is mostly used in transportation, urban planning, population health, and wellness \cite{Kellstedt2020ASR}. To drive the change in the mobility habits of people and to promote sustainable modes of transportation, an important role is played by the design and implementation of instruments that drive changes in individual behaviour.
For example, collaboration with formal and social groups like sport clubs has proven to be effective in providing significant strategies for motivating the level of participation in programs related to behavioural change toward active and sustainable mobility\footnote{To ensure that our transport systems meet society’s economic, social and environmental needs whilst minimising their undesirable impacts on the economy, society and the environment \cite{aei21235}.}\cite{HAUFE201649}. Practical guidance has been proven to be a significant inspiration and motivation for increasing the level of cycling, walking, and using public transport. 
Providing the target audience with the right kind of information such as the display of site-based data, and the promotion and organization of workshops to inform and train people on alternative and greener ways of travelling can significantly influence the mode or the choice of the mean of transportation \cite{Koszowski2019}.
Furthermore, accessibility instruments such as the availability of timetables, and the use of advertisements and leaflets can significantly promote and encourage the public to adopt modes of active and sustainable transportation \cite{Pajares2021}.

To promote active and sustainable mobility it is critical to analyze the effects of the decisions taken in support of this type of mobility and to evaluate the users' experience.
This can be achieved by monitoring the number of registered participants, identifying and profiling the most regular and active individuals, assessing the trips made and the modes of the trip, and analyzing their possible observed variation \cite{Pisoni2022}. 
Additionally, there is a significant need for systems that assess the levels of individuals who shift towards sustainable and active modes of transportation \cite{Koszowski2019,Pajares2021}.
Furthermore, activities like the modal split data, and the analysis of vehicles and people in terms of the saved vehicle kilometres, while a certain action is being performed, have also been observed to be a satisfactory and acceptable method of survey and evaluation \cite{Koszowski2019,Pajares2021}.

In this paper, we present an end-to-end solution, called Play\&Go Corporate, for enabling urban cyclability and its concrete exploitation in the realization of a home–to–work sustainable mobility campaign (i.e., \project) targeting employees of public and private companies. Play\&Go Corporate provides  a web console\footnote{\url{http://admin.playngo.it}} for the company mobility managers (MMs) to manage the necessary information (entity data, participating employees) and to visualize the information (trips/valid kilometers) of their employees. The web console allows each company to configure and manage all the information related to their employees participating in the running campaigns. Moreover, Play\&Go Corporate provides the Play\&Go Mobile App\footnote{\url{https://apps.apple.com/it/app/play-go/id1641103495}}$^{,}$\footnote{\url{https://play.google.com/store/apps/details?id=it.dslab.playgo}} for the employees, which allows them to track their home--to--work trips and to visualize the achieved results.

Mixing software solutions, economic incentives, data collection strategies and data analysis methodologies, we build a novel innovative, sustainable, and targeted framework to guide individuals toward a behavioural change. 
With the data collected through the user experience questionnaire, we show that by launching specific mobility campaigns, that exploit the Play\&Go Corporate solution, this significantly increased the use of bicycles for commuting from home to work.
With the data collected we analyze the bicycles' trajectories to understand how much a user travelled path diverges from the optimal one (i.e., shortest paths). Moreover, we assess how the street network’s safety level plays a role in the path selection of cyclists. 
Through the reported experience, we demonstrate how the solution proposed enables the understanding of the level of urban cyclability and helps the mobility (city) managers to derive the needed urban planning interventions.

Starting from the background and motivations that led to the realization of our solution (see Section \ref{sec:background}), we present the objectives of the \project mobility campaign we launched and the various steps that have been performed to engage both companies and employees (see Section \ref{sec:bike2work}). We then continue by providing details on Play\&Go Corporate, the technical implementation supporting its management and operations (see Section \ref{sec:software}). Section \ref{sec:data} shows how we support cities and local authorities to use the data coming from users' experience and behaviour to identify weaknesses in the city they manage. We conclude the paper with the qualitative (see Section \ref{sec:qualitative}) and quantitative (see Section \ref{sec:quantitative}) analysis of the \project experiments and with some conclusions and future work (see Section \ref{sec:conclusion}).

\section{Background and Motivations}
\label{sec:background}
\subsection{From Walkability to Urban Cyclability}
\textit{Urban cyclability} evolved from the notion of \textit{walkability}, expanding the study to include all the active modes \cite{Julian2020}. Cities that are ecologically sustainable, athletic, and socially viable rely heavily on bicycles \cite{Nielsen2018}. By combining the concepts of \textit{walkability} and \textit{urban cyclability}, the idea of \textit{likeability} was formed to assess the degree to which riding a bicycle is made easier \cite{Castanon2021}. For governments and other stakeholders interested in supporting \textit{sustainability goals in urban mobility} \cite{Wahlgren2014,Winters2016,Stewart2014}, the development of urban mobility policies and standards has become a crucial subject of inquiry and action \cite{Julian2020,Hagen2021}. With transportation accounting for nearly a third of energy usage in both the European Union (EU) and the United States (US), and ``single-occupancy vehicle (SOV)" daily commute rates even now high, there are numerous barriers to reducing pollution and other personal, societal, and ecological costs involved with this mode of transportation \cite{Castanon2021}. Although there was little interest in the issue before the start of the decade, there have been various projects since the turn of the century to assess the quality of surroundings for active modes \cite{Julian2020,Nielsen2013}. Nonetheless, the great majority of them focus only on walking \cite{Kang2019}.
Instead, an essential element of new development agendas for future cities is based on the need to adopt and advance sustainable mobility strategies \cite{vandecasteele2019future}. Such policies need to focus on making mobility affordable, accessible, and sustainable by enhancing walking, cycling and public transport services while reducing the impact of vehicular traffic at local and national levels.

\subsection{Active Mobility Advantages}
Research by the European Environmental Agency (EEA) shows that the transport sector is responsible for around a quarter of total greenhouse gas emissions in the EU (EEA, 2020\footnote{\urlx{https://www.eea.europa.eu/data-and-maps/indicators/transport-emissions-of-greenhouse-gases/transport-emissions-of-greenhouse-gases-12}}). A recent report on the latest trends declared that current levels are unlikely to achieve 2030 targets\footnote{\urlx{https://www.eea.europa.eu/data-and-maps/indicators/transport-emissions-of-greenhouse-gases-7/assessment}}, reinforcing the need to promote passive forms of emission reductions by reducing the need to rely on energy-intensive modes of travel.

The number of cars in Europe has increased by more than 10\% in less than ten years\footnote{\url{https://www.acea.be/statistics/tag/category/report-vehicles-in-use}}, with strong growth of traffic jams and rising CO2 emissions. To reduce these emissions and meet climate goals we need to drastically increase our efforts to reduce transport. As shown by previous research\footnote{\url{ http://stars-h2020.eu/wp-content/uploads/2020/04/STARS-5.1.pdf}}, \textit{active} and \textit{shared} mobility have proven to be a very relevant part of the solution. Bike sharing encourages citizens to cycle more or to start cycling again. Shared bike rides are often combined with public transport trips and offer an alternative for car use or even car ownership. The societal advantages of active and shared mobility are manifold: fewer car trips, more use of active and public transport, less CO2 emissions and pollution, and more valuable public spaces. There are also important individual advantages: people save time not having to look for parking, they are less frustrated not having to bother about administrative car issues, they live healthier through more active travels, and they can save a large amount of money (avoiding taxes, insurances, maintenance, parking fees). Nevertheless, the number of users embracing active mobility is still low: people are used to have their own car and it is hard to convince them to consider getting rid of it. The mental shift that fosters a behavioural change that eventually leads to a modal shift poses a tough challenge for both city planners and active mobility operators.

\subsection{Encouraging cyclability}
Due to the need for significant changes in individual behaviour and policy-making, mobility is expected to be a crucial area of focus in light of its impact on the environment, climate, and land use. These changes are likely to have long-term effects beyond the current generation and require a paradigm shift.

Thus, incentivizing citizens to be more engaged in sustainable mobility actions \cite{ROSE2007351} and changes is of paramount relevance to the success of the different \textit{Green Deal} \cite{greenDeal2021} aspects because mobility behaviour, like many other behaviours, has a strong impact on the environment.
Civic engagement that promotes ecological mobility can take different forms and may involve different populations. Each change may collectively bring a greater result. Technological solutions may facilitate some collective behavioural changes. However, it should not be forgotten that these changes start from individual changes \cite{Muller2015,anagnostopoulou_mobility_2020}. To achieve successful results, it is crucial to take into account all individuals in an inclusive approach. Sustainable mobility solutions have to be addressed according to the specific characteristics of the individuals, the institutions, the territories and the transport organizations.

To encourage cyclability, social characteristics can be viewed as a potential benefit particularly when it comes to inspiring change \cite{Julian2020,Castanon2021,Krenn2015,Winters2016}. Several persuasive principles that emerge around the social dimension have been established in the setting of behaviour change toward urban mobility \cite{Hagen2021}. Social comparison, social facilitation, normative impact, social learning, competitiveness, and praise are examples of these concepts \cite{Galanis2014}.

\subsection{Quantify Urban Cyclability}
Many researchers have developed indexes to assess specific aspects of the built environment that influence cycling behaviour and thoroughly \textit{quantify urban cyclability}, namely the extent to which an environment is friendly for bicycling \cite{Julian2020,Krenn2015,Schmid2021,mekuria2012low}. 
Previous research dealing with the development of \textit{urban cyclability indices} had to deal with challenges such as the time-consuming data gathering procedures \cite{Hagen2021,Schmid2021,Yeboah2014}, the balance between subjective and objective reasoning, the extraction of street-level data, and the standardization of spatial granularity \cite{Nielsen2018}. According to likeability literature studies, the historical progress of urban cyclability evaluation has indeed been driven by inventive applications of sophisticated new technology (i.e., street view imagery (SVI) \cite{Julian2020,Krenn2015}, computer vision (CV) \cite{Castanon2021}).

\subsection{Intelligent Transportation Systems and Cyclists' Behaviour}
Novel data sources can be employed to measure different aspects related to bike adoption and usage (e.g., demand prediction \cite{luca2021survey}). In this work, we focus on the work aiming at characterizing the behaviour of people on the move and strategies for street network developments. 
There are multiple works that aim at modeling the route choices of cyclists. Based on GPS and mobile phone data, researchers investigated aspects like statistically significant differences between optimal and chosen routes in different cities like Hamilton - Ontario, Canada \cite{lu2018understanding}, Portland - United States \cite{broach2012cyclists}, Waterloo region - Canada \cite{casello2014modeling}, Atlanta - United States \cite{misra2018modeling}, Phoenix - United States \cite{howard2001cycling}, and Zurich - Switzerland \cite{menghini2010route}. Other researchers explored the relationship between the chosen path and aspects like air pollution \cite{jarjour2013cyclist}, urban form \cite{raford2007space} and socio-demographic factors \cite{misra2018modeling}. Differently from the ones just mentioned, other works focused on using alternative data sources to inform municipalities and policymakers on optimal strategies to grow bicycle-related street networks. For example, \cite{szell2022growing} used OpenStreetMap data to estimate topological limitations of existing bicycle street networks to prioritize the development of new bike lanes. In \cite{vybornova2022automated}, the authors integrated mobility flows to make a more accurate decision based on gaps in the bicycle network and the usage of specific streets. Finally, \cite{folco2022data} highlighted how features related to demand and safety can be considered for prioritizing street network developments for e-scooters and bikes.

\section{Bike2Work: Objectives, Features and Management}
\label{sec:bike2work}
Within the AIR-BREAK project\footnote{\url{https://airbreakferrara.net/}}, behaviour change and awareness raising campaigns have the aim to inform citizens’ and raise their awareness on the possibilities and advantages offered by the available sustainable mobility services and to encourage the adoption of different, more sustainable, mobility habits. The Municipality of Ferrara, through the signing of a ``Memorandum of Understanding" with the Region Emilia-Romagna, promoter and financier of the initiative, supports the \project intended for public or private companies based in the Municipality of Ferrara. The project is part of the initiatives of sustainable mobility put in place to meet the new challenges of the COVID-19 emergency and to promote new strategies for sustainable transportation in order to curb the negative impact of air pollutant emissions, and it aims to promote the use of bicycles for home-work trips by providing an economic incentive to employees. Needs related to social distancing have, in fact, imposed a drastic downsizing of public transportation capacity making particularly relevant the encouragement of the use of bicycles and other modes of private transportation with low environmental impact.
With the \project initiative, the goal is to create greater awareness around this issue by promoting a more sustainable home-work mobility, contributing to the reduction of CO2 emissions.

Moreover, \project has the goal to support the company mobility manager in the promotion of sustainable mobility and transport demand management by analyzing the problems, needs and habits of workers, trying to orient them towards new habits of sustainable transport.

\color{black}
\begin{figure*}[ht]
\vspace{-0.2cm}
\centering
\includegraphics[width=.8\textwidth]{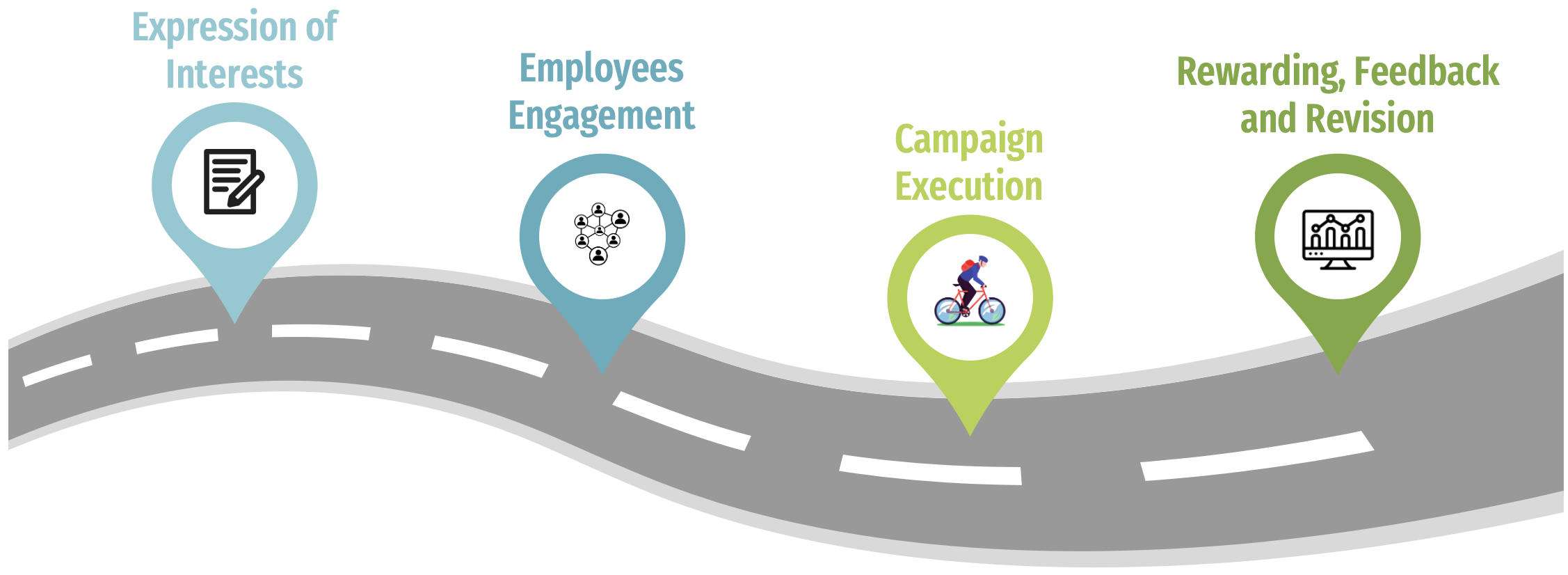}
\caption{\project workflow.}
 \label{fig:workflow}
  \vspace{-0.1cm}
\end{figure*}

The adoption of technological solutions alone cannot make transport more sustainable; to do so it is necessary to involve people and guide them towards a behavioural change. To achieve these goals, \project~ intends to engage companies with their employees to build new innovative, sustainable, and targeted solutions that can improve quality of life more effectively.
 
The specific objectives of this initiative are:
\begin{itemize}
    \item To support workers in switching to sustainable mobility habits resulting in reduced CO2 emissions;
    \item To support public/private companies in the adoption of policies, initiatives, and the development of urban mobility plans;
    \item To increase the perception of corporate (ecological) Social Responsibility and improve Total Quality Management (TQM) within companies;
    \item To increase cooperation between different modes of transport and promote interconnection and interoperability between existing transport networks;
    \item To increase the attractiveness of sustainable transport modes through the implementation of different measures such as proposing new private mobility policies, promoting public transport, and pooling and sharing services.
\end{itemize}

The city of Ferrara\footnote{\url{https://www.comune.fe.it/}} is a medium-sized Italian city located between Bologna and Venice, along the Po river, with an overall number of inhabitants of 131,000 distributed in an area of 400 $Km^{2}$. The city has broad streets and numerous palaces dating from the Renaissance, when it hosted the court of the House of Este. Moreover, Ferrara is a pretty flat city where weather conditions are never particularly impactful. 

The \project initiative was launched in May 2021 and it is still running. It is part of the sustainable mobility initiatives put in place from the Municipality of Ferrara to promote the use of bicycles for home-to-work trips by providing an economic incentive to employees of public or private companies.

\project~  provides incentives for sustainable mobility through an economic contribution for workers committed to using bicycles for home-to-work trips. Employees of participating companies are rewarded for their home-to-work trips by bike with economic incentives in their paychecks (0.20 € per Km, max 50 € per month, max 20 km per day). Mobility managers and employees are supported by Play\&Go Corporate, a software platform and mobile app described in detail in Section IV, and the overall campaign participation is supported by a specific workflow depicted in Figure \ref{fig:workflow}.

Each interested company provides, through an \textbf{Expression of Interests}, the following information: (a) all the company data, (b) the list of mobility managers (MMs) with their related information, (c) the details of the headquarters participating in the \project campaign with the declaration of closure days (e.g., holidays). 

As soon as a company has been involved, it has the role to \textbf{engage} their employees. Each employee that agrees to participate is registered at the Play\&Go Corporate console with all the needed information. In this phase, the MM sends an email to each participating employee. In this email each employee receives:

\begin{itemize}
    \item The presentation of the campaign with the relative regulations and information regarding data processing and privacy;
    \item The instructions to perform the registration to the campaign and to download the software application needed to participate. 
\end{itemize}

As soon as an employee accepts the invitation to participate to the \project campaign and the registration is done, s/he can start tracking the home--to--work and work--to--home bike journeys (\textbf{Campaign Execution}). 

The workflow terminates leveraging the data collected via Play\&Go Corporate to give \textbf{feedbacks} to the municipality and policy makers and to inform the MMs about their employee performance and using the data to allocate the needed payroll \textbf{rewards}. In this phase, to assess the employee experience and receive feedbacks useful for the initiative \textbf{revision}, a user experience questionnaire is planned.
Since the duration of the \project initiative is three years, there are three iterations of the workflow. Companies that are already enrolled continue without interruption with the possibility of add/deleting employees. For the new companies, there is a window where they can express their interest in joining (January-March).

Even if the workflow depicted in Figure \ref{fig:workflow} has been defined and implemented to manage the \project campaign in Ferrara, it has been defined in such a way that it can be reused in different contexts (different cities, countries, etc.) and for different sustainable mobility objectives (different sustainable transportation means such as public transports, different rewarding mechanisms). This is possible thanks to the software system we have implemented that we introduce in the next Section. Play\&Go Corporate has been implemented with the goal to be general-purpose (for different cities and with different supported transportation means) and flexible enough to be extended and customized to meet the needs of the specific mobility campaign.

\section{The Play\&Go Corporate Software System}
\label{sec:software}
The workflow and all the features presented in the previous Section have been used to guide the implementation of the Play\&Go Corporate solution. It is an innovative software system that provides a console of data, information, recommendations, and simulations for MM to assess, also through what-if analysis, the environmental impact of employees' commuting, to evaluate changes because of specific measures and actions, and to plan optimal and sustainable workers' mobility strategies.
To achieve the identified objectives, Play\&Go Corporate provides: 
\begin{itemize}
    \item A \textbf{web console}\footnote{\url{http://admin.playngo.it}}  - for the MM of the company to manage the necessary information (entity data, participating employees) and to visualize the information (trips/valid kilometers) of the employees.
    \item The \textbf{Ferrara Play\&Go Mobile App}\footnote{\urlx{https://play.google.com/store/apps/details?id=it.smartcommunitylab.playgoferrara},}$^{,}$\footnote{\url{https://apps.apple.com/us/app/id1526145980}} - for the employees, which allows them to track their home--to--work trips and to visualize the achieved results.
\end{itemize}

The functionalities supported by the Ferrara Play\&Go Mobile App concern the employee's registration and the management of the employee's profile, the tracking of sustainable trips, the inspection of employee's results (e.g., points earned, badges and badge collections, active challenges with completion status, weekly and global leader boards ranking, personal mobility diary), information on weekly and global prizes, as well as the access to mobility campaign rules and regulation. The application provides a homepage, in which a summary of the employee's state is presented. The homepage also presents a set of frequent and immediate actions that the user can perform, e.g., trips tracking. 

\begin{figure}[htb]

\centering
\includegraphics[width=.6\linewidth]{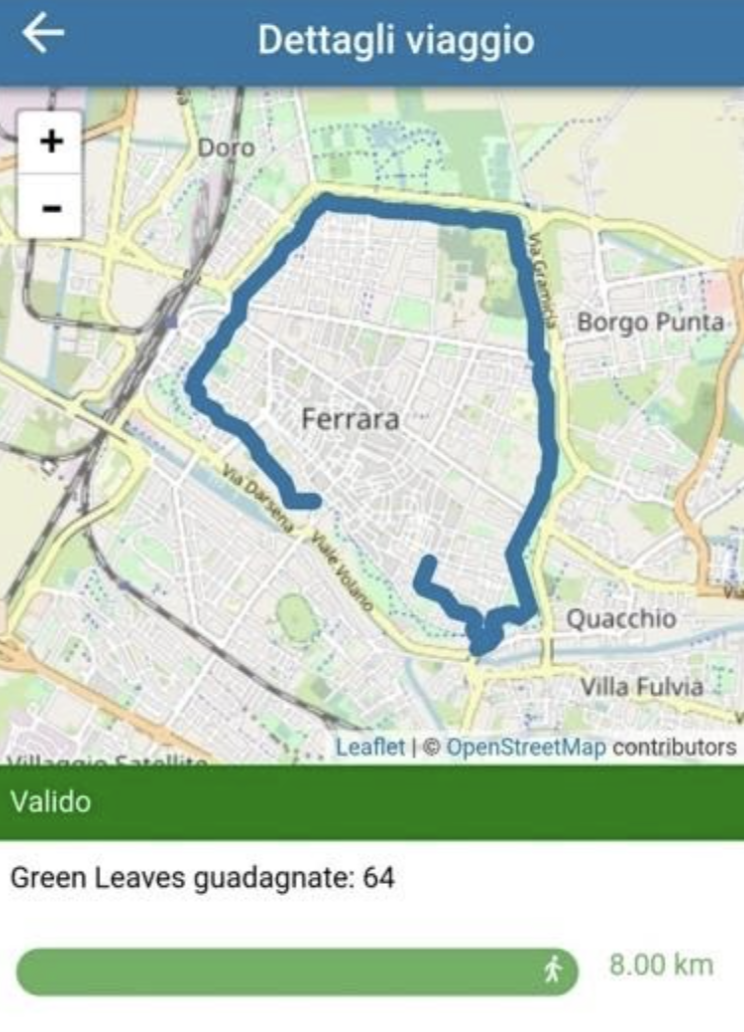}
\caption{Trip tracking view.}
 \label{fig:tracking}
\end{figure}

In the \project campaign,  employees can track trips by bike and can visualize the trips on a real-world map (see Figure \ref{fig:tracking}), both in real-time while they are recording them during their journey, and for past trips stored in their profile.

\begin{figure}[htb]
\vspace{-0.2cm}
\centering
\includegraphics[width=.4\textwidth]{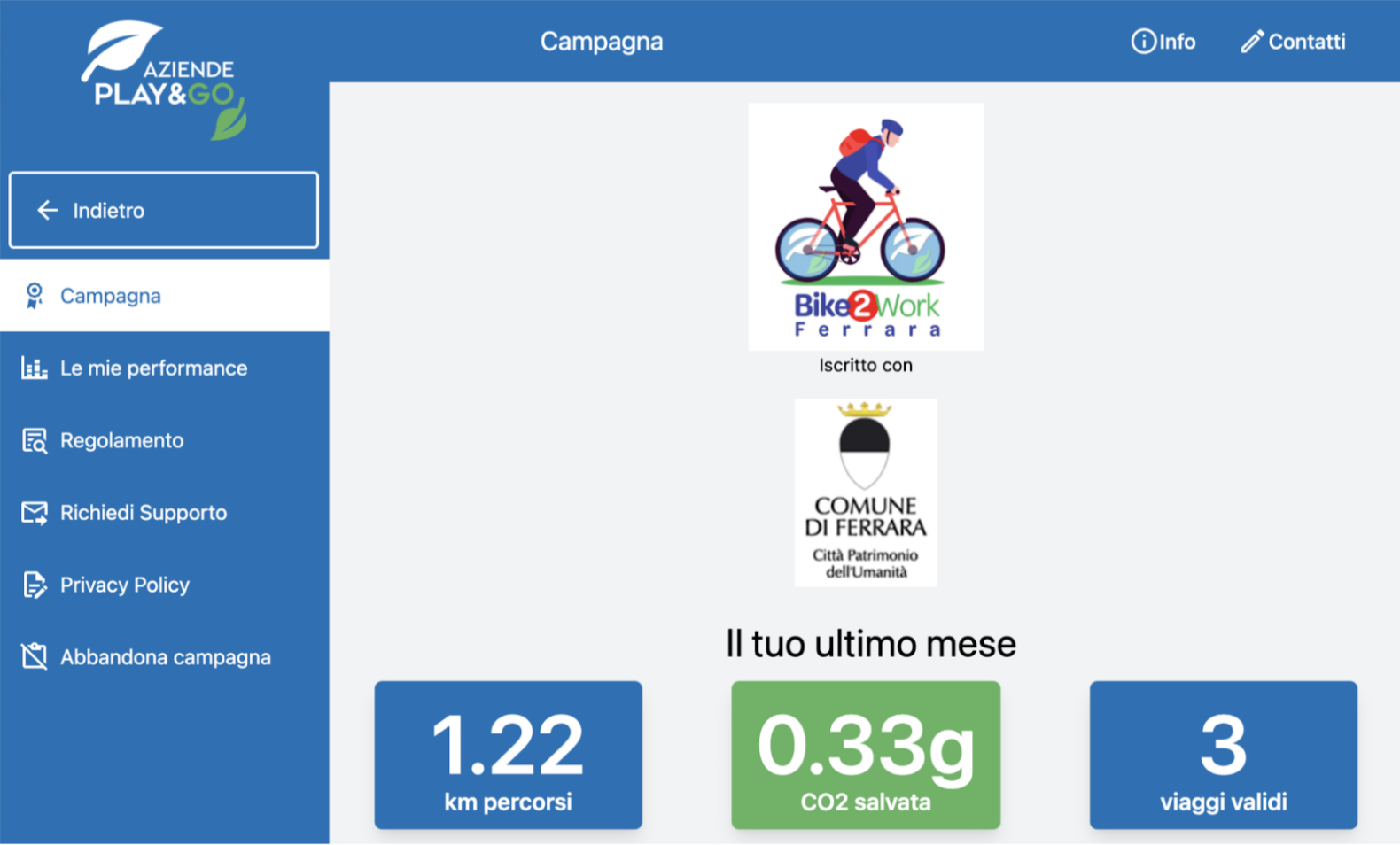}
\caption{Employee dedicated area.}
 \label{fig:performance}
  \vspace{-0.1cm}
\end{figure}

Each employee who has joined the \project campaign can directly enter in a dedicated area (see Figure \ref{fig:performance}) where she/he can monitor her/his progress in the campaign. Access to this private area can take place directly through the Ferrara Play\&Go App through a dedicated web link\footnote{\url{https://aziende.playngo.it/}}. The main objective of this area is to show to each employee her/his behaviour regarding home--to--work mobility. For this reason, it is possible to visualize dedicated information on the Km traveled, the CO2 saved and the number of valid trips. Moreover, each employee can consult the \project~ campaign regulations, the privacy information document, and any news dedicated to the campaign in execution.

To validate the bike journeys done to reach the work locations by the employees, the mobile app exploits a dedicated \textit{trip validation component}. It is a \textit{mode detection} component that allows comparing a user’s “actual” mode, based on traces of the user’s position and activity sampled during her/his trip, vs. her/his “declared” mode, that is, the mode selected by the user when starting the journey tracking through the app. Only valid trips are sent as player actions to the Gamification Engine \cite{KazhamiakinLMS21} and contribute to the progress of the player in the game. In any case, the trips that are not valid for the \project campaign are still displayed in the user app and possibly counted for other active campaigns that consider other types of sustainable transportation means (i.e., city campaign that considers buses, walk, and train as valid means).

The mode detection algorithm can be configured, depending on the application setting, to also consider some additional information to  ``certify" the tracked data. For example, in the case of the \project campaign, the employees are assigned to a specific set of company headquarters that they could reach every working day. The trip validation component checks if each single journey starts or arrives from/to one of the declared locations and if the trip is performed within a company's working day. If the trip validation component considers the trip valid, the corresponding employee action is sent to the \textit{gamification engine component} that updates the employee's state correspondingly. Otherwise, the \textit{trip validation component} provides a specific motivation for not considering the trip valid (e.g., too fast). The validity outcome, in case of a valid trip, or the explanation of the specific motivation, in case of an invalid trip, is presented to the employee in the mobile app.

\begin{figure*}[htb]
\centering
\includegraphics[width=\textwidth]{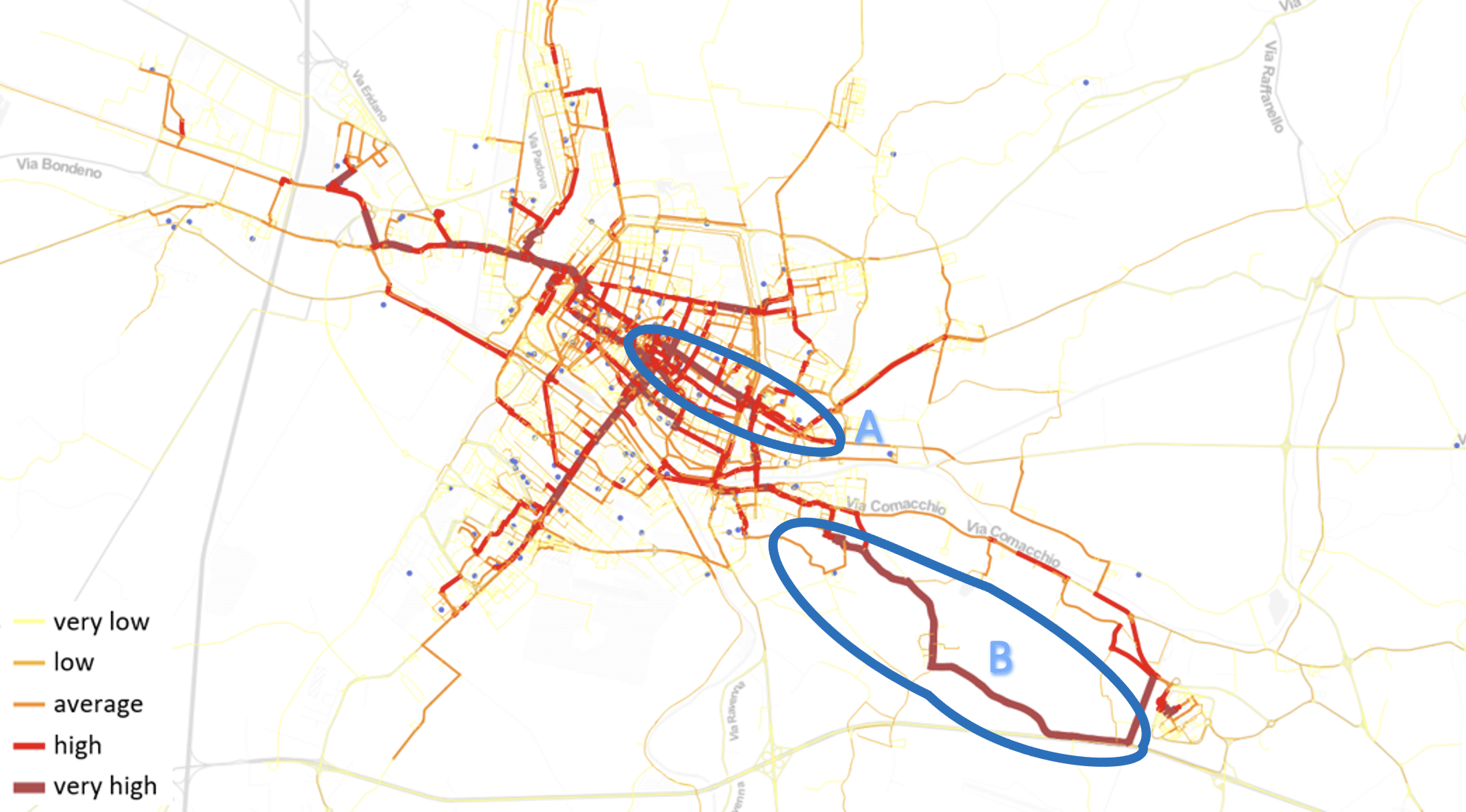}
\caption{Map of \project~commuters in Ferrara (at December 2021).}
 \label{fig:map3}
\end{figure*}\textbf{}

\subsection{Open Data Analysis Support}
\label{sec:data}
Mobility data are everywhere, but not always accessible and understandable. Sometimes cities and local authorities do not know how to get and use them efficiently and how to convey analytics and derived information in a simple way to a non-specialized audience. To reach this goal, we need multidisciplinary approaches that combine information design and data science techniques.
A key element is represented by the availability of open data representing transport networks: indeed, open data (i.e., datasets publicly available with open license like OpenStreetMap\footnote{\url{https://www.openstreetmap.org/}}) are a real treasure that too often remains untapped. Among others, open data about transport networks (road networks, public transport networks, inland water networks, infrastructures for bikers and pedestrians, etc.) represent key datasets, geospatial by nature \cite{EU2020}. 

In the Play\&Go Corporate system, the impact of the running initiatives are analysed using anonymized data from “Ferrara Play\&Go” mobile app (i.e., raw GPS points), followed by a map-matching algorithm that reconstructs each trip from the raw GPS logs using the OpenStreetMap (OSM) road network.
Generally speaking, a map-matching algorithm \cite{Quddus2007} is an automatized procedure that combines measures from one or more positioning devices with data from a road network map to provide an enhanced positioning output. This task is usually not straightforward because of the combined effect of measurement errors in positioning data and accuracy errors in road network data. 
In the context of the \project campaign in Ferrara, the exploited map-matching procedure handles the positioning uncertainties adopting a Bayesian approach of Maximum Likelihood: the data are projected on the road segments that have the higher probabilities of having generated them \cite{GiovanniniPhD}.
The overall procedure can be divided in different phases. Before the actual map-matching of GPS trajectories takes place, some initialization operations are performed to speed up the following elaborations: road network data for the area are loaded in memory and a road proximity map is created. This proximity map allows for a fast identification of the road arcs that are close to every given spatial position inside the area. Once the initialization step is completed, the map-matching can start. First of all, the data from each bike trip goes through a trajectory aggregation stage, that serves the purpose of removing useless data and aggregating useful GPS data into trajectories. 
Then, GPS trajectories are processed in sequence through the two last steps of the procedure: (i) the projection of GPS data into the surrounding road elements and (ii) the identification of the optimal path between the projected data. 
Another set of automatic procedures calculates different indicators at a single road segment, by timestamp. These procedures are designed to provide practical and easy answers to typical use cases:
\begin{itemize}
    \item What are the most used routes within the city? 
    \item Do they match infrastructures for bikers? 
    \item What are the critical points for cyclists’ safety? 
    \item Where are cyclists riding the wrong way?
\end{itemize}

To showcase the results, different map applications have been deployed for sharing data. Web maps have been implemented based on a set of open-source Javascript libraries (OpenLayers\footnote{\url{https://openlayers.org/}}) for displaying spatial data in web browsers, with simple-but-effective interactive maps showing where the streets mostly used by the ‘BIKE2WORKers’ are in Ferrara. Different spatio--temporal indicators have been developed. The map in Figure \ref{fig:map3} shows for example where are the streets mostly used by the \project participants in Ferrara, from May 2021 until end of December 2021.
In the map, the two major findings are highlighted in blue colour:
\begin{itemize}
    \item  Corso Giovecca, which cuts the city center from east to west and which in the western part is lacking dedicated cycle lanes despite being very popular (see the blue ellipse with label A in Figure \ref{fig:map3}).
    \item The new cycle lane, opened in early 2021, that leads from the center to the hospital of Ferrara in Cona village (located to the east) and which appears to be widely used by commuters working at the local health authority and at the University (see the blue ellipse with label B in Figure \ref{fig:map3}).
\end{itemize}

In order to make the results of this initiative continuously accessible, an interactive map has also been made available online\footnote{\urlx{https://sit.comune.fe.it/allegati/mappe/airbreak/MobilitaAirFriendly.html}}. The map can be browsed (zoom/pan) and queried. By clicking on a street segment, a user gets information about the number of transits in the selected segment. The number of transits can be filtered for a specific day of the week or other time periods (e.g., weekend, entire week, month, etc..). The selected data can be exploited by the city administrators to identify weaknesses in the network they manage in a secure and timely manner \cite{EU2020}. Additionally, they can be efficiently used to create innovative services, like on-demand services, ride-hailing, etc. so to allow public and private transport companies to increase their performances and enlarge their offerings. In the case of \project, OSM road network is the main source for running algorithms to calculate which are the streets mostly used by the \project participants in Ferrara and what are the critical points for cyclists’ safety.

\section{Participants' Experience and Behaviour Change}
\label{sec:qualitative}
To carefully analyze the user experience and behaviour change related to the \project campaign, we developed a questionnaire, which was sent to all the participants.

\subsection{Questionnaire definition}
\label{sec:questionnaire}
We developed an ad-hoc user experience questionnaire\footnote{The full questionnaire in the original Italian language with the English translation opposite can be retrieved here:  \url{https://osf.io/8t4bm}} to gather useful and necessary information related to the \project campaign, in order to improve the campaign features before the next runs, and at the same time gather information about the user experience, with items related to the adopted rewarding system, the experienced fun, and the behaviour change towards eco-sustainable behaviours, which was the ultimate goal of the \project campaign. Moreover, comments about the campaign’s strengths and weaknesses, and suggestions for possible implementations were collected.

In the case of the most commonly used Likert-type scales (5-point and 7-point scales), the edges are excessively distant from the center (i.e., from strongly disagree to strongly agree). This common practice may lead participants to be biased to select items closer to the center rather than the distant ones \cite{alexandrov2010characteristics,samuelson1988status}. Also, verbal anchors tend to influence the perceived distance between the points of the scale to which they refer \cite{lantz2013equidistance} and this is the basis for a perceived asymmetry in the scale that can also cause an \textit{end of scale} effect. This asymmetry in the scale perception can be stemmed changing the verbal anchors or removing the central neutral item.
Neutral points are non-differentiating because neutrality may be interpreted by raters as ambiguity, ambivalence,  irrelevance, or indifference \cite{armstrong1987midpoint}. 
Hence, we decided to present the items aimed at analyzing the user experience with a 6-point Likert-type scale \cite{batterton2017likert} (from 1 = very negative, to 6 = very positive), and tailoring the items on the bike use according to the following criteria: "Never/almost ever", "Less than once a week", "1/2 times a week", "3/4 times a week", "Almost always/always"\footnote{The translation has been made by the authors for this current paper. The original quotes are "Mai/quasi mai", "Meno di una volta a settimana", "1/2 volte a settimana", 3/4 volte a settimana", "Quasi sempre/sempre".} . 

The questionnaire is divided into theoretically tailored different parts that collect information on different aspects of the user experience: (1) the overall experience (i.e. "\textit{How would you rate your experience with Ferrara Bike2Work?}"), (2) users' habits ("\textit{How often did you use your bike to commute from home to work before participating in Ferrara Bike2Work?}", "\textit{How often did you use the bike for commuting from home to work during the Ferrara Bike2Work initiative?}"), (3) satisfaction related to several features of \project (such as tracking, economic incentives, accession procedure, and data visualization; i.e. "\textit{How would you rate your experience with respect to using the Ferrara Play\&Go App for travel tracking?}", "\textit{Did you appreciate the economic incentives associated with the initiative?}", "\textit{How would you rate your experience with the process of joining the initiative?}", "\textit{How would you rate your experience with respect to using the Play\&Go Corporate Web App to view your results?}"), and (4) the intention to participate in future campaigns (i.e. "\textit{Would you participate in a new edition of FerraraBike2Work?}").
Then, the questionnaire provided other ad-hoc items useful for the analysis of users' behavior: (1) the amount of bike use before and during the initiative, and (2) the distance from home to work. This part allows quantifying the extent to which \project had a role in the selection of the means of transport.
Finally, the questionnaire offered several open questions for evaluating the strong and weak points of \project, and collecting suggestions and useful data for improving future \project campaigns.

\subsection{Methods}
\label{sec:methods}
Exploiting the in-app notification system supported by the Play\&Go Corporate solution, the questionnaire link was sent to all active participants (538 employees from 55 companies) during the \project campaign in December 2021, after 7 months of the campaign (it started in May 2021). There were no rewards for completing the questionnaire. According to the general data protection regulation (GDPR), the data collected were anonymous, hence no user's personal information has been collected. A total of 162 users' answers were used for the questionnaire analysis.

\subsection{Results}
\label{sec:questionnaire_results}
\begin{figure*}[htb]
\centering
\includegraphics[width=.9\textwidth]{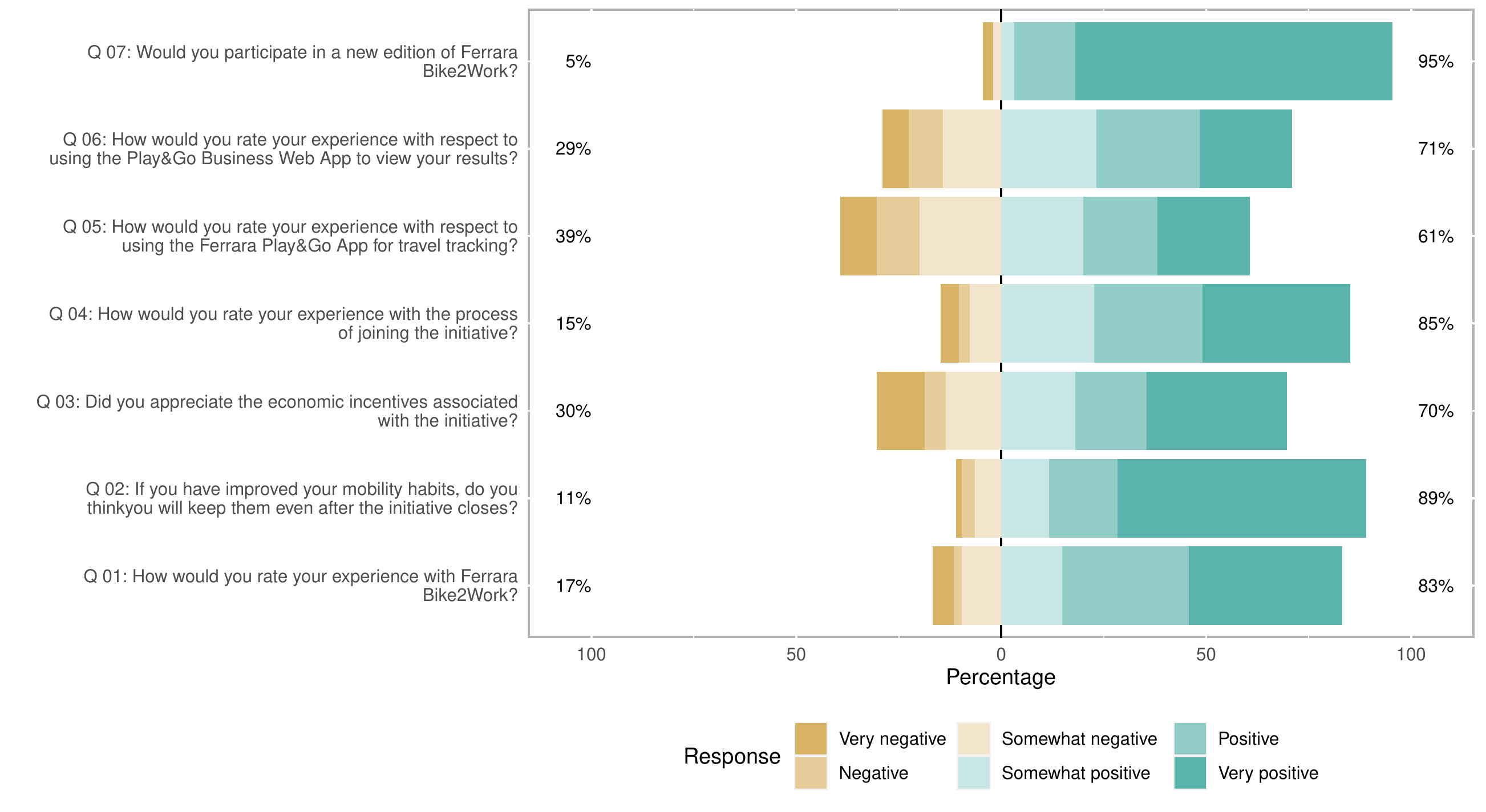}
\caption{Results to the Likert-type items of the questionnaire}
 \label{fig:questionnaire}
\end{figure*}

Overall, the Likert-type items analysis showed an extremely positive value for the items related to the overall experience with the initiative (83\% of users; mean = 4.7 $\pm$ 1.44), intention to keep the habits adopted during \project (89\% of users; mean = 5.21 $\pm$ 1.2), the intention to participate again in the initiative (95\% of users; mean = 5.6 $\pm$ .98), and the experience with the process of joining the initiative (85\% of users; mean = 4.66 $\pm$ 1.4). The overall result with respect to the other features measured by the questionnaire always showed positive results, but smaller in magnitude. 
In fact, the results related to the Play\&Go Corporate Web App to monitor the campaign results showed the 71\% of very positive values (mean = 4.17 $\pm$ 1.48), the user experience related to the app feature for the travel tracking reports a 61\% of positive values ( mean = 3.9 $\pm$ 1.6), and the experience related to economic incentives showed a 70\% of appreciation (mean = 4.2 $\pm$ 1.7).
Figure \ref{fig:questionnaire} shows an overview of the responses to the Likert-type items of the questionnaire.

\textbf{}

Moreover, according to some users' quotes (e.g., ``[I put a low evaluation level for the economics incentives] \textit{because not being able to trace the paths, and I did not have any economic incentive}"), we decided to analyze the causal relationship between the three features reported as least positive and the overall experience evaluation.

The data analysis reported a positive correlation between the overall experience and all the three \project features: the perception of the tracking ($R^{2}$ = .54, p $<$ .001), the evaluation of economics incentives ($R^{2}$ = .1843, p $<$ .001), and the results visualization through the web app ($R^{2}$ = .417, p $<$ .001).

\begin{figure}[t]
\vspace{-0.2cm}
\centering
\includegraphics[width=.45\textwidth]{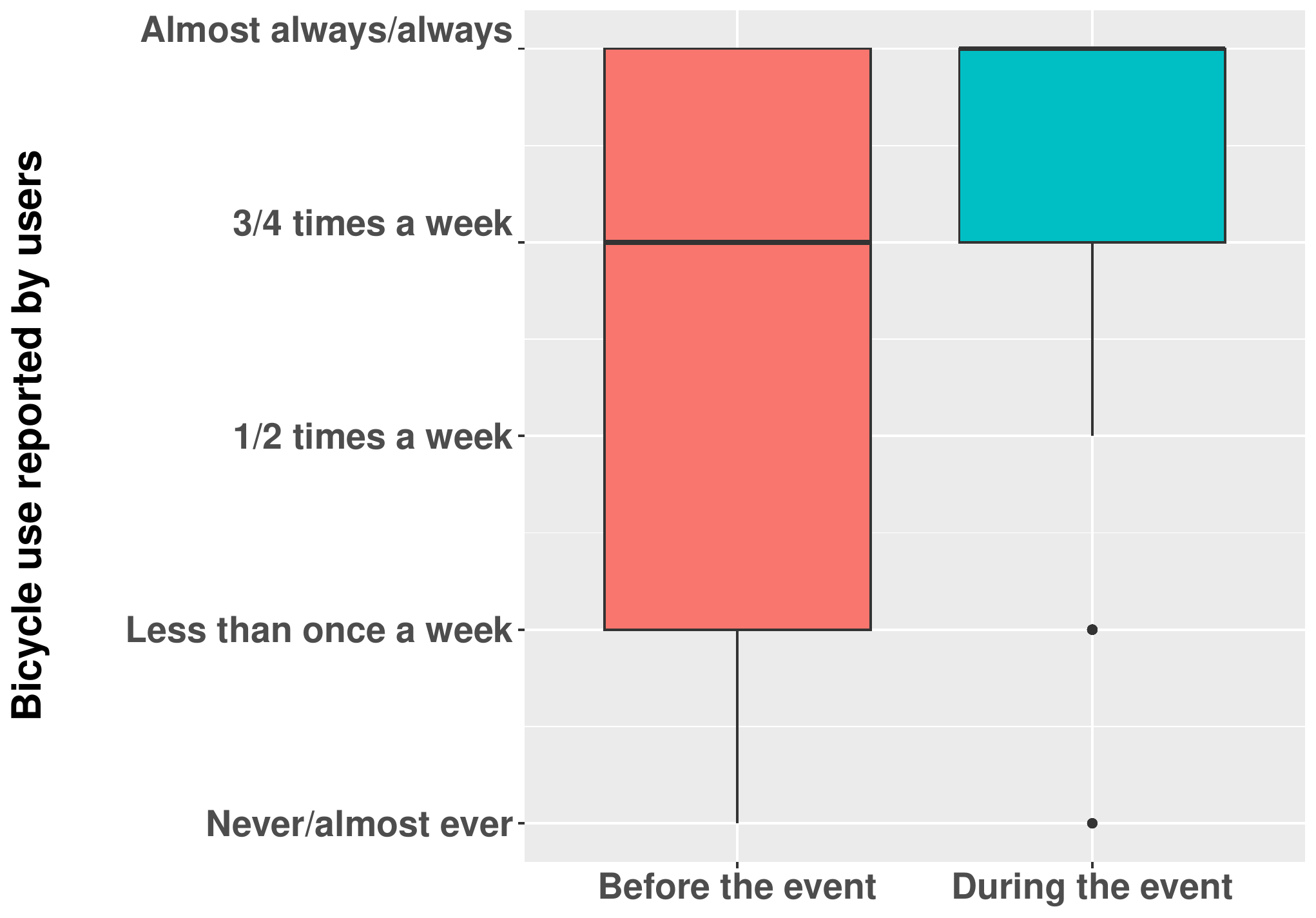}
\caption{Use of bicycle before and during the \project campaign.}
 \label{fig:pre-post}
  \vspace{-0.1cm}
\end{figure}

To analyze whether there was a role of the initiative in the use of bicycles, we first determined the data distribution by performing a Shapiro-Wilk test \cite{thode2002testing}. It showed that the distribution of the use of bicycles before the initiative departed significantly from normality (W = .707, p $<$ .001). Based on this outcome, we run a Mann-Whitney-U test \cite{nachar2008mann}, finding a statistically significant difference in the use of bicycles before and during the initiative (W = 16,245,  p $<$ .001, $\delta$ = -.24), suggesting that the participation in \project associated users means on transportation selection for commuting from home to work (see Figure \ref{fig:pre-post}).

\textcolor{black}{
We decided to investigate also the role of the means of transportation on the user experience.
Interestingly, the analyses didn't find differences in the overall experience (W = 2,661, p $>$ .05), and in the intention to participate again in the initiative (W = 3,231, p $>$ .05) according to the means of transportation used before the \project campaign.}

Then, we decided to analyze eventual differences in the overall experience according to the kilometers driven each day by users. For this analysis, we considered only three factors: ``1-5 KM" (n = 95), ``5-10 KM" (n = 52), and ``10-30KM" (n = 12), since the two factors ``Less than 1 KM" and ``More than 30 KM" were reported by one user each.
The analysis didn't show any significant difference according to the distance traveled daily from home to work ($R^{2}$ = -.01, p $>$ .05).

\subsection{Discussion}
During the \project campaign, users reported an increase in bicycle use, bringing the percentage of participants using the bicycle for home-to-work commuting at least three times a week from 55.8\% to 77.9\%, with an increase of 40\%. The experience has been positive for the users, regardless of the means of transportation used before the initiative. Nevertheless, some elements need to be monitored during the upcoming \project campaigns (i.e., tracking, visualization, and economic incentives).

The results also suggest that the \project campaign was perceived equally positively by (1) the users who continued to use the bike to move, and (2) the users who changed their means of transportation during the campaign regardless of the distance traveled.
However, due to a lack of related data, it was not possible to comprehensively evaluate the overall experience with the distance traveled by users, and the different means of transportation used before the campaign.
We expect to investigate in depth these relations in future \project campaigns.

Moreover, some issues in the tracking feature would seem to have led to a decrease in economic incentives perception, and consequently to a lower overall experience evaluation. We expect that by improving the tracking of routes, we could (1) increase the receipt of incentives, (2) improve their evaluation, and (3) lead to a better perception of the \project campaign.

\section{Use Case: Leveraging Play\&Go Data for Informing Municipalities}
\label{sec:quantitative}

\begin{figure*}[t]
    \centering
    \includegraphics[width=\textwidth]{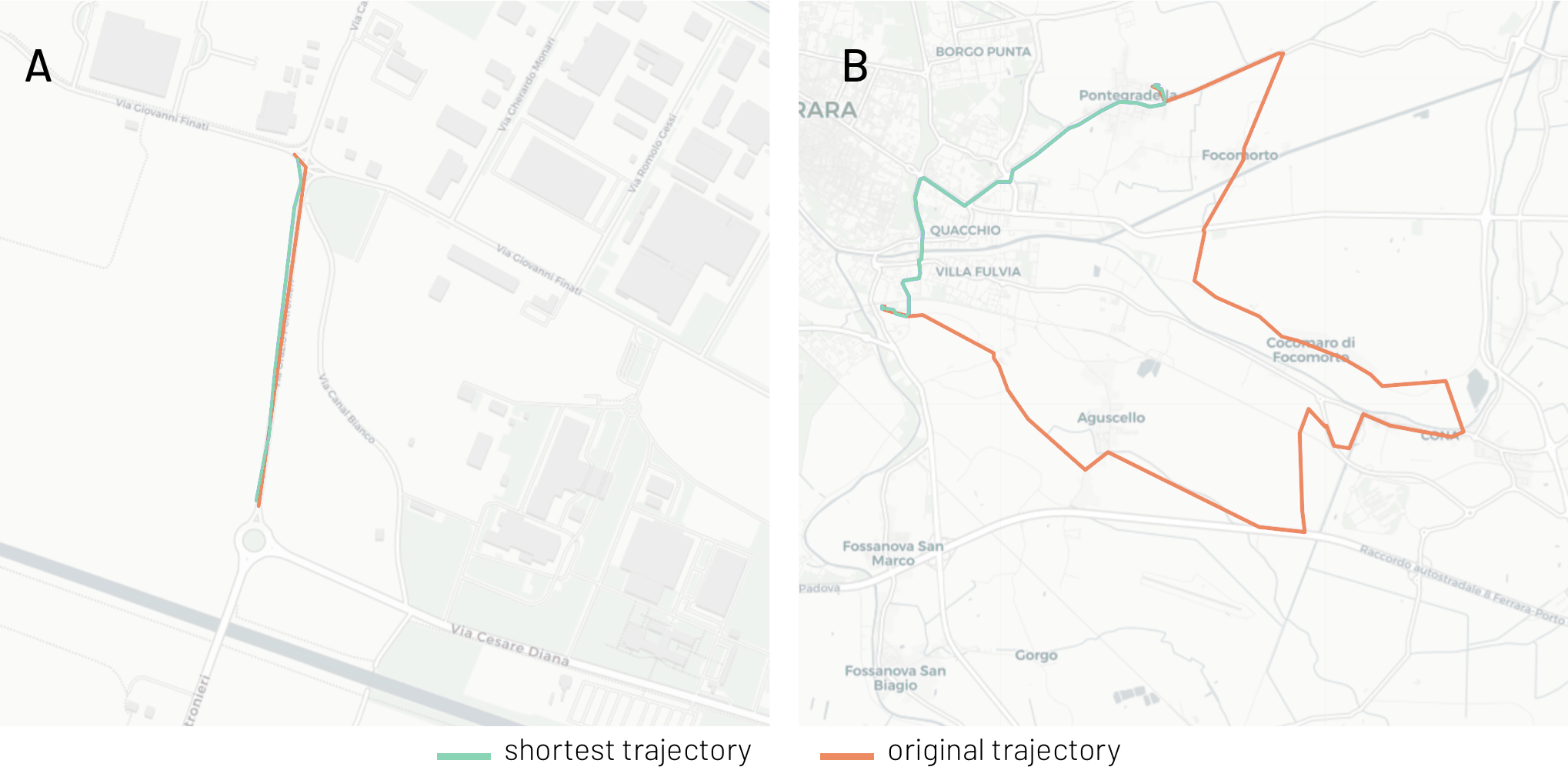}
    \caption{Users' trajectories example. (A) A user's short trajectory difference between the original trajectory and the shortest one; (B) a long trajectory difference between the original trajectory and the shortest one.}
    \label{fig:fig1}
\end{figure*}
 The scope of this Section is to show a way to leverage data collected via Play\&Go Corporate to inform municipalities and policymakers to make data-driven decisions potentially.
While campaigns like Ferrara \project provide significant nudges and incentives to encourage citizens to use bicycles for home-work commuting, an important aspect that may hinder the regular use of bicycles may be represented by the lack of safe routes to ride. There are already some works discussing ways to optimize the development of bicycle networks. For example, \cite{szell2022growing} leverages network science algorithms to spot topological limitations of existing bicycle networks to prioritize the development of new bike lanes. In \cite{vybornova2022automated}, the authors integrated mobility flows to make a more accurate decision based on gaps in the bicycle network and the usage of specific streets. In our work, 
through the data collected by the Play\&Go app, we analyze the \project bicycles trajectories to understand how much the paths travelled by users diverge from the optimal one (i.e., shortest paths in terms of time and length). With users' mobility data and street-level information, we aim to understand whether the underline street network’s safety level plays a role in the paths selected by cyclists.
These analyses may provide insights and inform municipalities on how to identify streets that may need interventions to sustain and promote greener means of mobility.

To access street-level information, we download the street network from OpenStreetMap (OSM) keeping only the roads on which bikes are allowed (e.g., we remove highways), and then analyze and process the anonymized GPS traces collected during the \project campaign. The dataset includes 26,221 trajectories generated by 605 users over six months from May 2021 to September 2021. 

Formally, we define a trajectory as follows: 

\bigskip{ \begin{definition}[Trajectory]
A spatio-temporal point $p=(t, l)$ is a tuple where $t$ indicates a timestamp and $l$ a geographic location. A trajectory $P_i^u = p_1,p_2,\dots,p_n$ is a time-ordered sequence of $n$ spatio-temporal points visited by a user $u$, who may have several trajectories, $P_{1}^u, \dots, P_{k}^u$, where all the locations in $P_i^u$ are visited before locations in $P_{i+1}^u$.
\end{definition} }

\bigskip{In our dataset, $l$ is a tuple of latitude and longitude with the GPS points sampled every five seconds. The street network downloaded from OSM can be formalized as follows: }

\bigskip{\begin{definition}[Street Network]
A street network $SN = (V, E)$ is a directed graph where the vertices $v \in V$ are intersections or initial/final points of a road and the edges $e \in E$ are the streets. Each vertex $v_i$ has an associated latitude and longitude, while each edge $e_{(v_i, v_j)}$ has a set of properties $a_{e_{(v_i, v_j)}}$ (e.g., speed limit, size, etc.) and is connected with two vertices.
\end{definition} }

\bigskip{To} unveil potential issues on the road network, assuming that a study participant tries to reach their workplace/home as fast as possible, we evaluate how much a participant's observed trajectory deviates with respect to its corresponding shortest paths (both in term of time and length). For all users' trajectories, we compute the shortest path between trajectories' origins ($p_1 \in P$) and destinations ($p_n \in P$). First, we map the origins and the destinations with the nearest $v_i \in SN$ using the ball tree algorithm for Haversine nearest neighbour search implemented in \textit{osmnx}\footnote{https://osmnx.readthedocs.io}. 
Then, we apply the Dijkstra algorithm to compute the shortest path on $SN$. We generate two different shortest paths that we use to create two weighting schemes for the edges: (i) the \textit{time} needed to commute on an edge, and (ii) the \textit{length} (in meters) of the edge. The lower the time or the length, the more similar the observed trajectory is to the shortest one (see Figure~\ref{fig:fig1}).

A reason for cyclists to deviate from the optimal path may not just be related to the distance to commute. The street network may play an important role. Suppose that a cyclist has to commute between two destinations nearby, but the shortest path is a dangerous road. It is likely that a cyclist will use a longer but safer path. To validate this hypothesis, we computed the so-called Level of Traffic Stress (LTS) \cite{mekuria2012low}, an index that, given some meta-information about the street (e.g., speed limit, street size, type of cycle lane), classifies the streets into four different levels of danger for cyclists. LTS 1 represents streets with no or little stress, suitable for children (e.g., a large cycle lane completely separated by other streets), while LTS 4 represents streets for expert cyclists which are more dangerous to travel (e.g., a road where speed limits for cars are significantly high and there are no dedicated cycle lanes). 

We compute the LTS for the city of Ferrara using the algorithm provided by BikeOttawa\footnote{https://github.com/BikeOttawa/stressmodel}. This algorithm takes into consideration the characteristics of a street provided by OSM and leverages them to compute the LTS score. After discarding the streets on which bikes are not permitted (e.g., highways), the model retrieves the information on the remaining streets such as the number of lanes, the maximum permitted speed, the presence of parking lots on the side of the street, and whether or not the bike line is separated from the rest of the street. 
A combination of the aforementioned data corresponds to the LTS score of a street. For example, the LTS is equal to 1 when there is a separated bike line. When the bike lane is adjacent to a road with a maximum speed greater than 65 km/h, the LTS score is 4.


Before computing an LTS score for each edge $e \in E \in SN$, we performed a map-matching process to infer the path traveled on the road network from a participant's GPS trajectory \cite{Yang2018FastMM}.

To compute the stress level of a bike trip we average the LTS scores of each of the road segments included in a trajectory. In particular, a trajectory $T$ is now a set of edges where each edge $e_i$, among the other attributes, have an associated LTS. We indicate the LTS of a specific edge as $e_i^{\text{LTS}}$. The average score for a trajectory is computed as follows:

$$
\text{score}(T) = \frac{\sum_{e \in T} e^{\text{LTS}}}{||T||}
$$

The score for a trajectory is the average LTS score of all the street segments used to travel from the origin to the destination.

\begin{figure}
    \centering
    \includegraphics[width=\columnwidth]{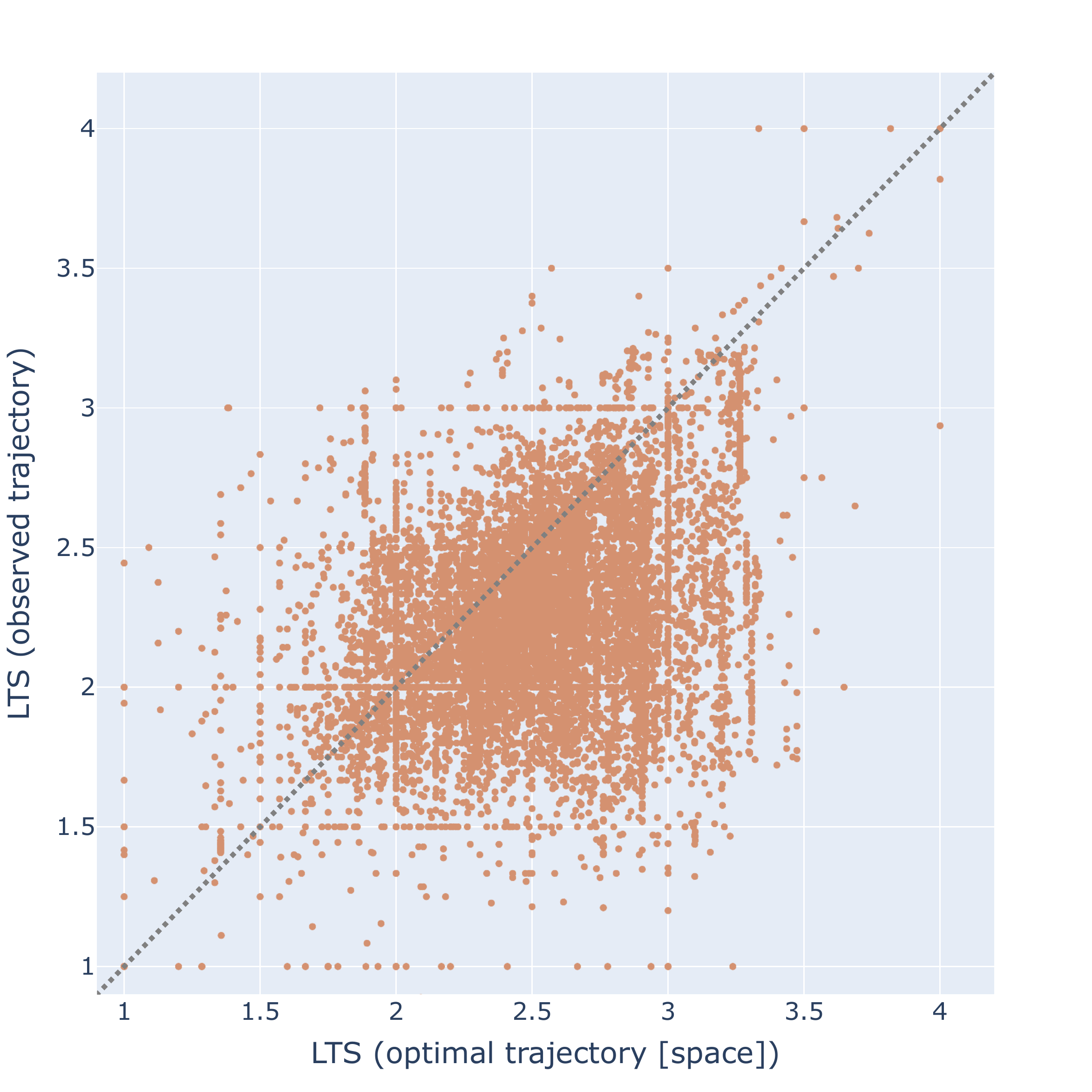}
    \caption{
    LTS score of observed trajectories (y-axis) and LTS score of optimal trajectories in terms of space (x-axis). Each point represents a trajectory. The diagonal gray dotted line represents the case in which both the shortest path and the observed paths have an equal LTS. The majority of points lie under the diagonal which means that cyclists tend to travel on roads with lower LTS (safer roads) despite not being the shortest ones.
    }
    \label{fig:fig2}
\end{figure}

Comparing the observed original trajectories with the corresponding optimal trajectories (i.e., shortest paths), we observed that cyclists in Ferrara not only tend to travel longer paths (original trajectories (median) $= 3451.82$ meters; optimal trajectories (median) $= 2948.81$ meters) but also travel on streets that display lower stress levels. 
Regarding the latter, in Figure~\ref{fig:fig2}, for each trajectory, we plot a point and we report the observed LTS score (on the y-axis) and the score associated with the relative shortest path (on the x-axis). Points on the diagonal (the grey dotted line) are the trajectories with an observed LTS score equal to one of the observed trajectories. Points below the diagonal are travels that have an observed score lower than the one of the optimal path. This means that the individual who registered the trajectory decided to take a safer path, even if longer (Figure \ref{fig:fig2})

\begin{figure}
    \centering
    \includegraphics[width=\columnwidth]{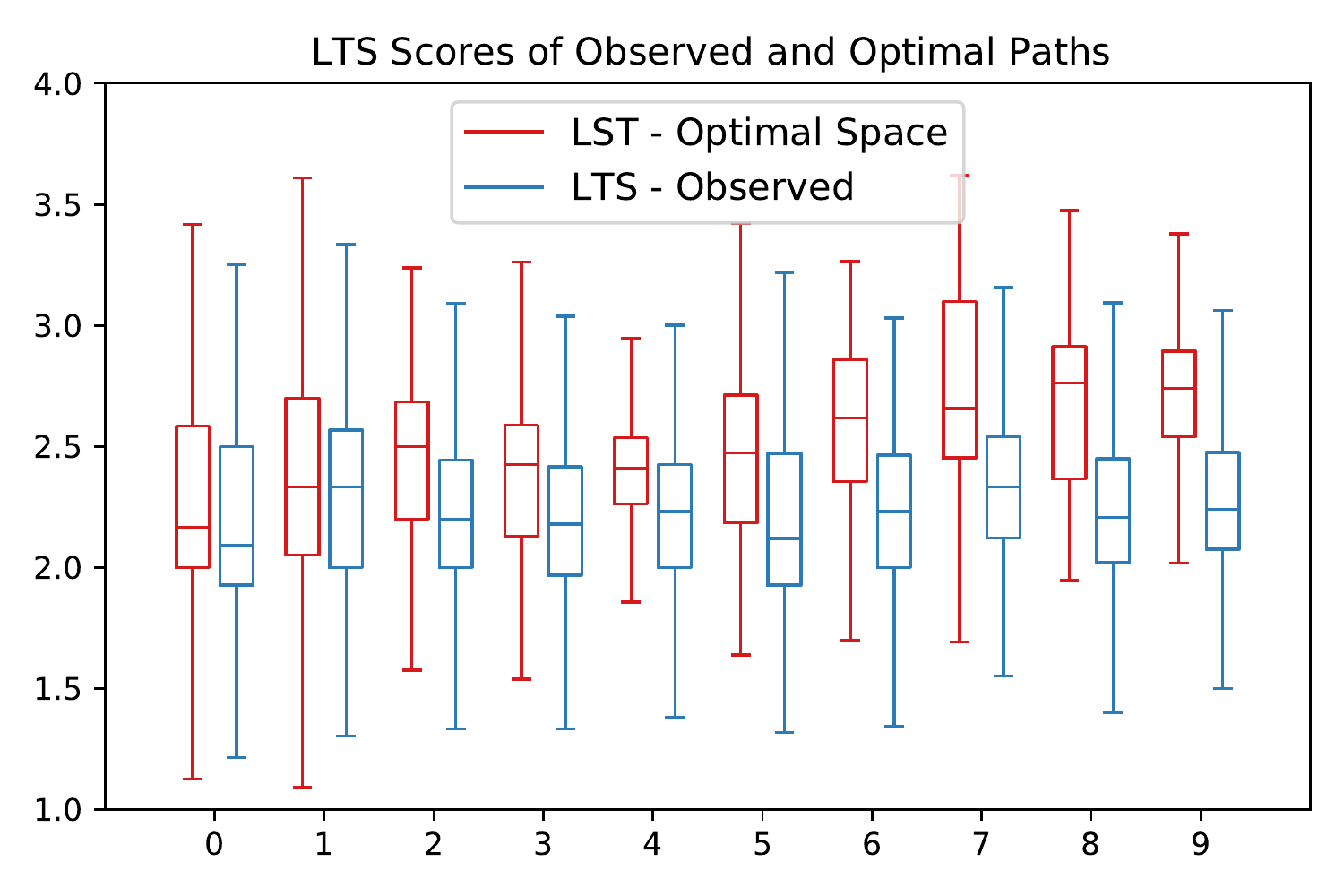}
    \caption{We divided the trajectories sorted by  growing distance between origin and destination into ten equal-sized bins. For each group, we estimate the median (line in each box) and the variance (box-tails) of the LTS scores observed (in blue) and one of the optimal paths (in red). It is possible to see how, in general, the LTS of the optimal path augments (i.e., more dangerous path) as the distance between origin and destination increases. On the other hand, the LST scores of the observed paths tend to be stable regardless the distances commuted.}
    \label{fig:fig3}
\end{figure}

Looking at Figure \ref{fig:fig2}, it is not clear if the safeness of the selected path is also related to the length of the travel. To answer this question, we consider all the travels shorter than 10 Km and we divided them into deciles based on the distance between the origins and destinations. In other terms, trajectories in the first bin are the shortest while trajectories in the last bin are the longer. For all the trajectories we computed the observed LTS and the LTS of the optimal path. In Figure \ref{fig:fig3}, we can see two boxes for each bin representing the median and standard deviation of the LTS scores of the trajectories in each bin. The red boxes are the scores of the optimal trajectories while the blue boxes are the scores of the observed trajectories. It clearly emerges that regardless of the distance traveled, on average bikers tend to travel on safer roads (observed LTS is lower than the optimal LTS). Moreover, we observe that the longer the distance traveled, the larger the difference between the optimal and observed LTS.

\subsection{Discussion}
All the analyses presented in this Section can be used by policymakers and municipalities to prioritize street renovation and/or construction in order to sustain greener mobility. Also, the data and insights can eventually be used to validate policies. 
For instance, the analyses presented could be potentially used to find the trajectories that show a considerable divergence in the average score between optimal and observed trajectories. Therefore, based not only on the stress level but also on how much a street is used by cyclists, we could locate the more problematic streets. This in turn could be used to inform municipalities and possibly guide them in improving the cycling road network. 
Finally, the continuous data collection through the app may help policymakers and municipalities to evaluate the effectiveness of the implemented policies (e.g., check whether or not the design of safer streets lead to changes in users' behaviours or a greater adoption of sustainable transportation means).

\section{Conclusions and Future Work}
\label{sec:conclusion}
Among the major impacts the \project initiative includes the analysis of open-source large data sets which supports policy and decision makers, urban planners and designers. As we have seen in this paper, Play\&Go Corporate, the end-to-end software solution presented, not only supports the Mobility Managers and the employees throughout the sustainable mobility campaign but has been conceived to understand the progress and the impact of the running initiative.  After the first \project campaign's execution, some initial results have been obtained and reported. We will continue running the campaign for the next year (till October 2023). Taking advantage of the experience gained and reported here, the next steps will be to improve both the methods and supporting technologies to revitalize the campaigns introducing new motivational methods (i.e., personalized gamification challenges to introduce competition among employees).

Play\&Go Corporate has been designed and developed to be used in any city and with any number of users. It is our goal to replicate the \project experience in other cities not only on the Italian territory but also abroad.
During the \project campaign, no socio-cultural or personal data were collected from users. According to some references in the literature \cite{zahedi2021gamification, oliveira2018exchange}, these components may have contributed to a different user perception of the campaign. Future directions for the \project campaign include also expanding to new realities, hence it is crucial to analyze these factors. We therefore expect to be able to use a more comprehensive questionnaire (which is under development) in the future. 
Moreover, this version of the campaign is devoid of gamified elements, which instead can be useful to foster behavioural change \cite{BASSANELLI2022103657}. Hence, we are working on adding gamification elements during the home-work commute to encourage healthy competition among users. For the realization of the gamified part, we will rely on contextual design frameworks \cite{bassanelli2022gamidoc}. Moreover, we plan to explore various modeling approaches that take into account latent factors that impact user behavior following an investigation and modeling process \cite{LaPaix2021}.

\section*{Acknowledgements}
This work is supported by the AIR-BREAK project funded through the ERDF Urban Innovation Actions 2020 UIA 05-177. We also acknowledge the support of the PNRR ICSC National Research Centre for High Performance Computing, Big Data and Quantum Computing (CN00000013), under the NRRP MUR program funded by the NextGenerationEU.

\bibliographystyle{IEEEtran}
\bibliography{biblio}
\newpage
\begin{IEEEbiography}[{\includegraphics[width=1in,height=1.25in,clip,keepaspectratio]{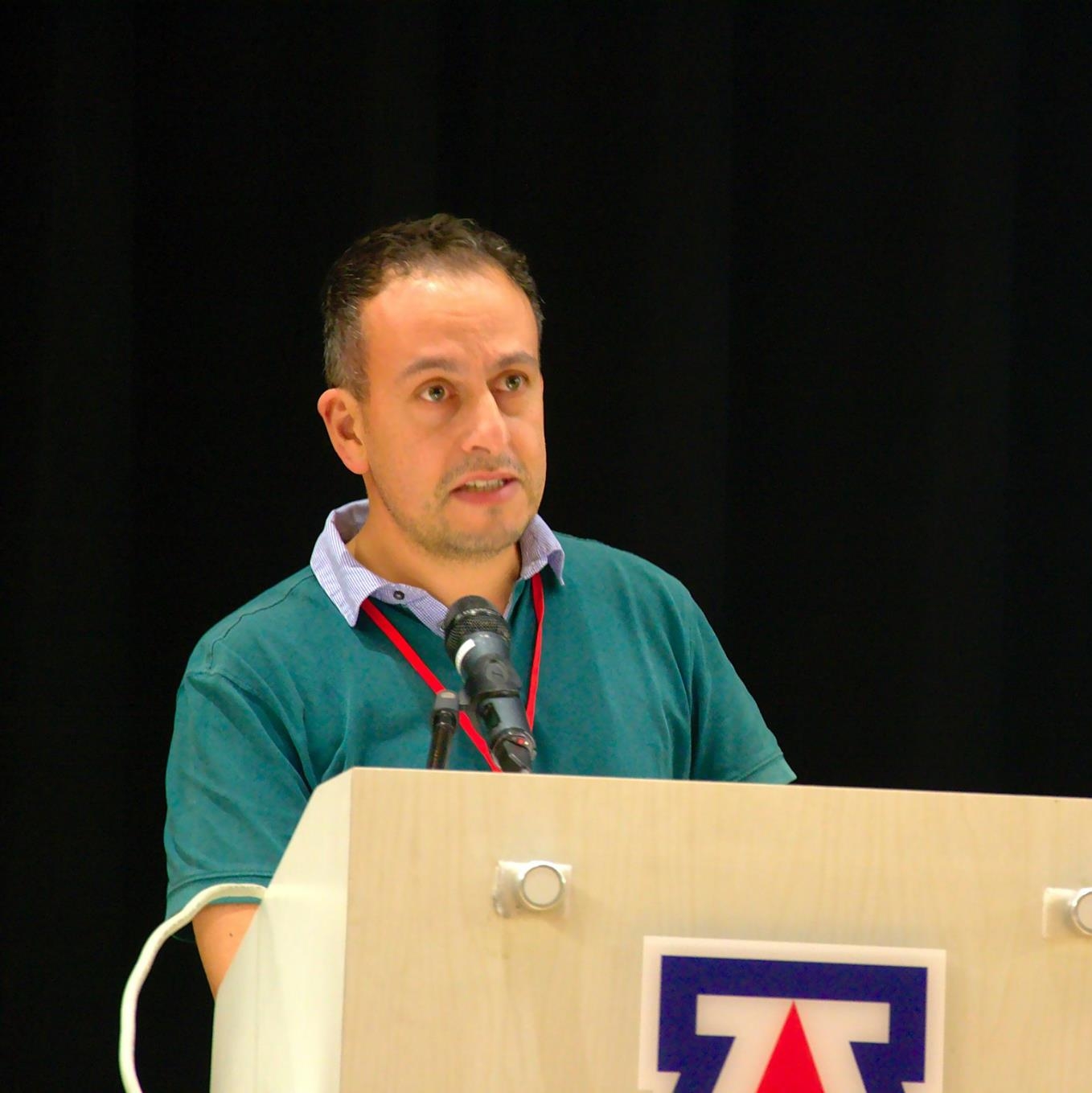}}]{Antonio Bucchiarone} is a Senior Researcher in the Motivational Digital Systems (MoDiS) unit in Fondazione Bruno Kessler (FBK), Trento, Italy.  His main research interests include: Self-Adaptive (Collective) Systems, Domain Specific Languages for Socio-Technical Systems, Smart Mobility and Multi-Agent based modeling and simulation.
He has been actively involved in various European research projects in the field of Self-Adaptive Systems, Smart Mobility and Service-Oriented Computing. He was the General Chair of the 12th IEEE International Conference on Self-Adaptive and Self Organizing Systems (SASO 2018). He is an Associate Editor of the IEEE Transactions on Intelligent Transportation Systems (T-ITS), the IEEE Software, and IEEE Technology and Society Magazine. You can contact the author at \url{bucchiarone@fbk.eu}  or visit \url{https://bucchiarone.bitbucket.io/}.
\end{IEEEbiography}
\vspace{-12 mm}
\begin{IEEEbiography}[{\includegraphics[width=1in,height=1.25in,clip,keepaspectratio]{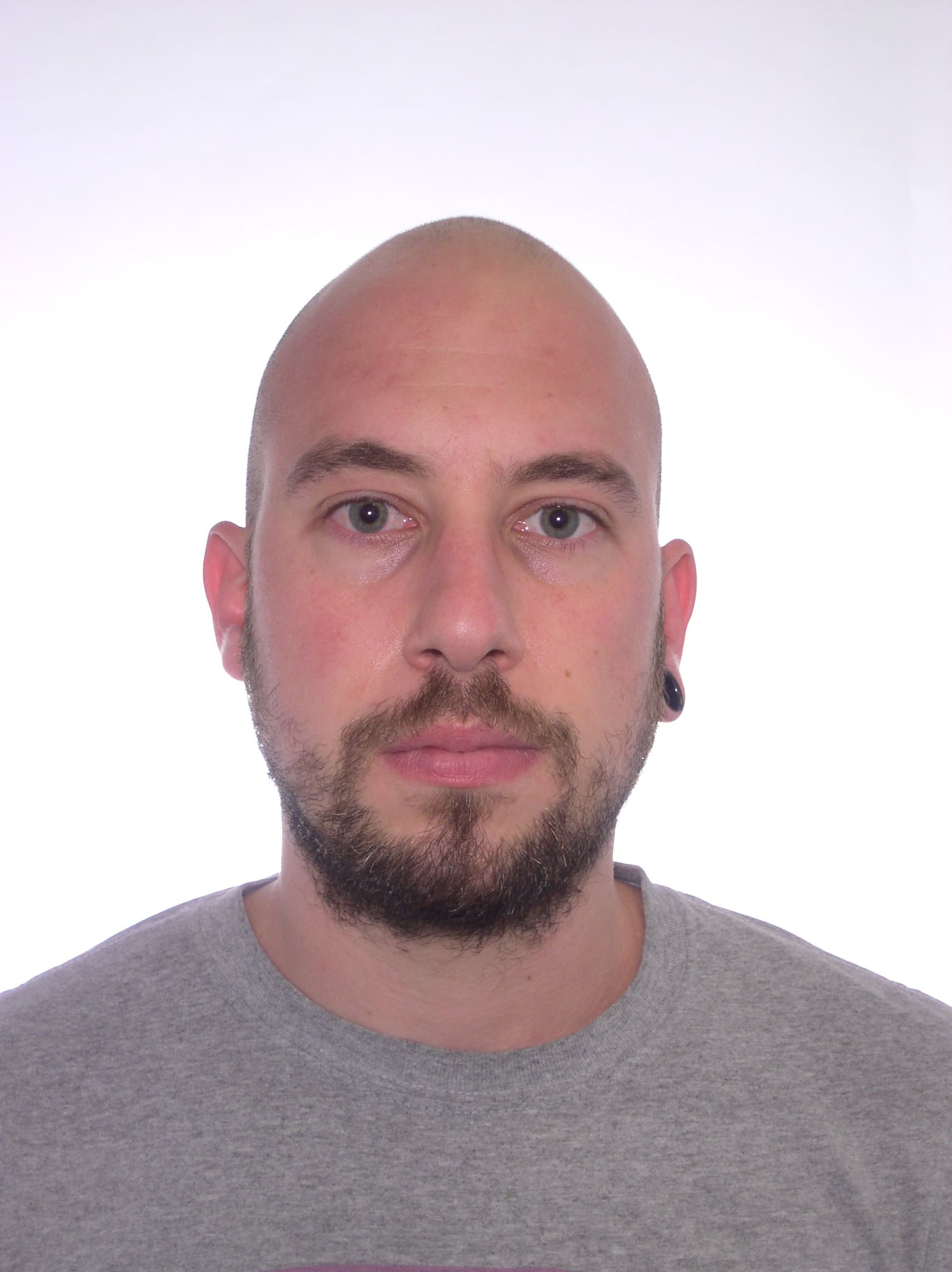}}]{Simone Bassanelli} is a Ph.D. student at the University of Trento in the Faculty of Psychology and Cognitive Science and in the Motivational Digital System (MoDiS) unit in Fondazione Bruno Kessler (FBK).
His main research interests include gamification and gameful design techniques, in particular the analysis of a holistic and proper way to design, develop and evaluate gamified solutions. 
You can contact the author at \url{simone.bassanelli@unitn.it}.
\end{IEEEbiography}
\vspace{-12 mm}
\begin{IEEEbiography}[{\includegraphics[width=1in,height=1.25in,clip,keepaspectratio]{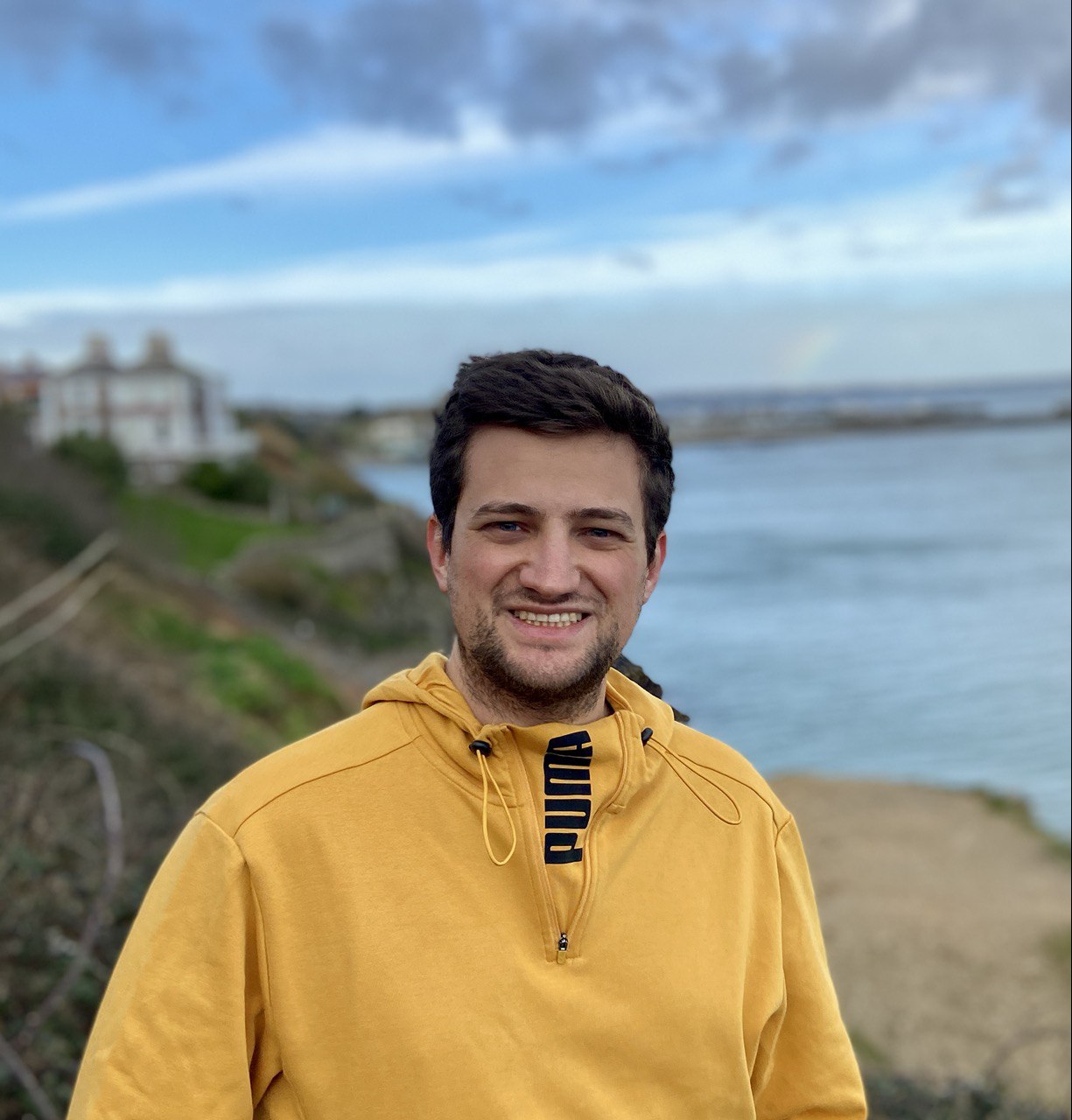}}]{Massimiliano Luca}  is a Ph.D. student in the Faculty of Computer Science at the Free University of Bolzano and in the Mobile and Social Computing Lab at Fondazione Bruno Kessler.  He is broadly interested in computational sustainability and machine learning methods to predict and generate human mobility. Previously, he was a research intern at Centro Ricerche Fiat (now Stellantis) and Telefonica Research.
\end{IEEEbiography}
\vspace{-8 mm}
\begin{IEEEbiography}[{\includegraphics[width=1in,height=1.25in,clip,keepaspectratio]{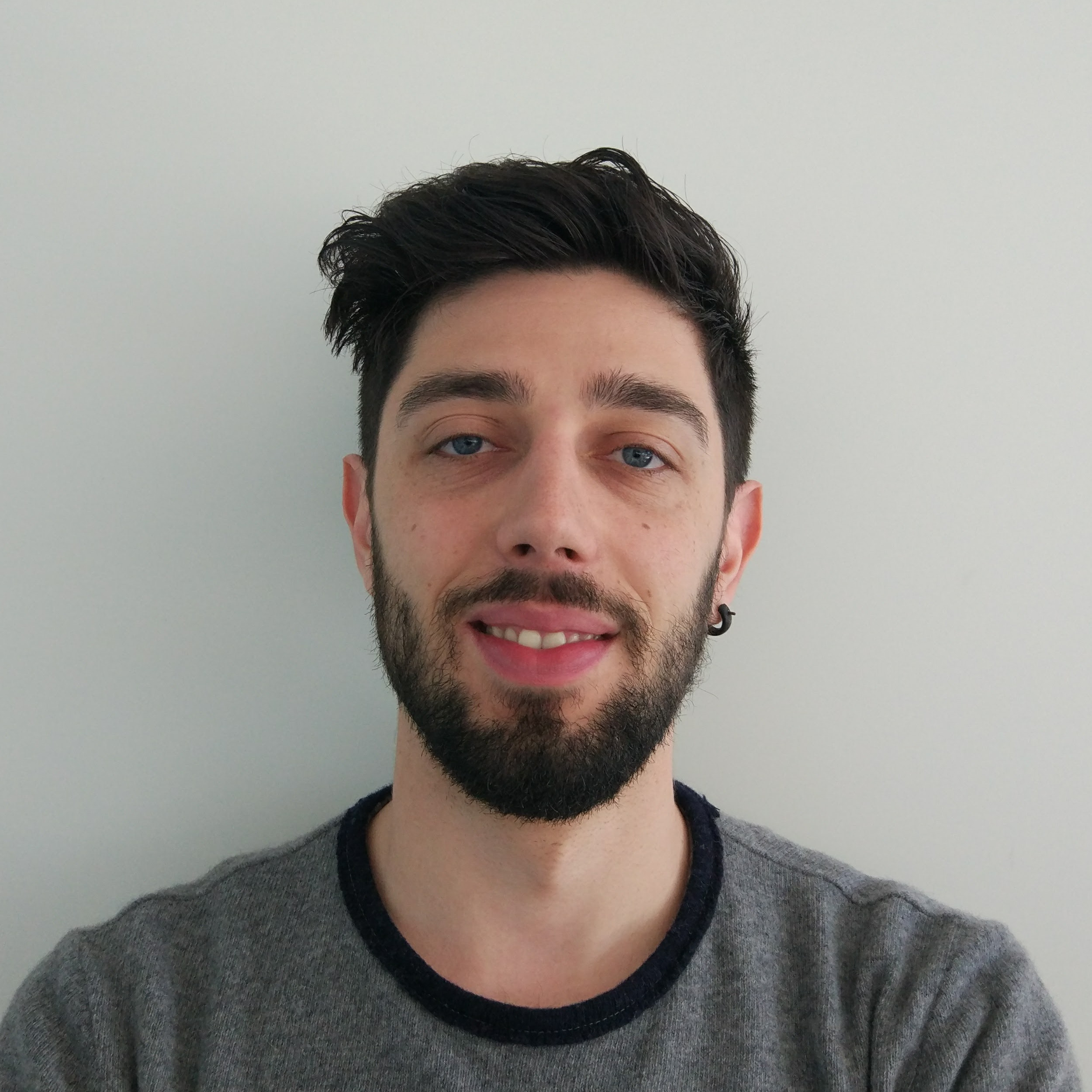}}]{Simone Centellegher} completed his PhD in Information and Communication Technology at the University of Trento, Italy, where he is currently a researcher in the Mobile and Social Computing Lab at Fondazione Bruno Kessler (FBK). His research interests focus on human behaviour understanding and in particular on the analysis of spending patterns from
credit card transactions data, and social interactions and mobility patterns extracted from mobile phone data.
\end{IEEEbiography}
\vspace{-8 mm}
\begin{IEEEbiography}[{\includegraphics[width=1in,height=1.25in,clip,keepaspectratio]{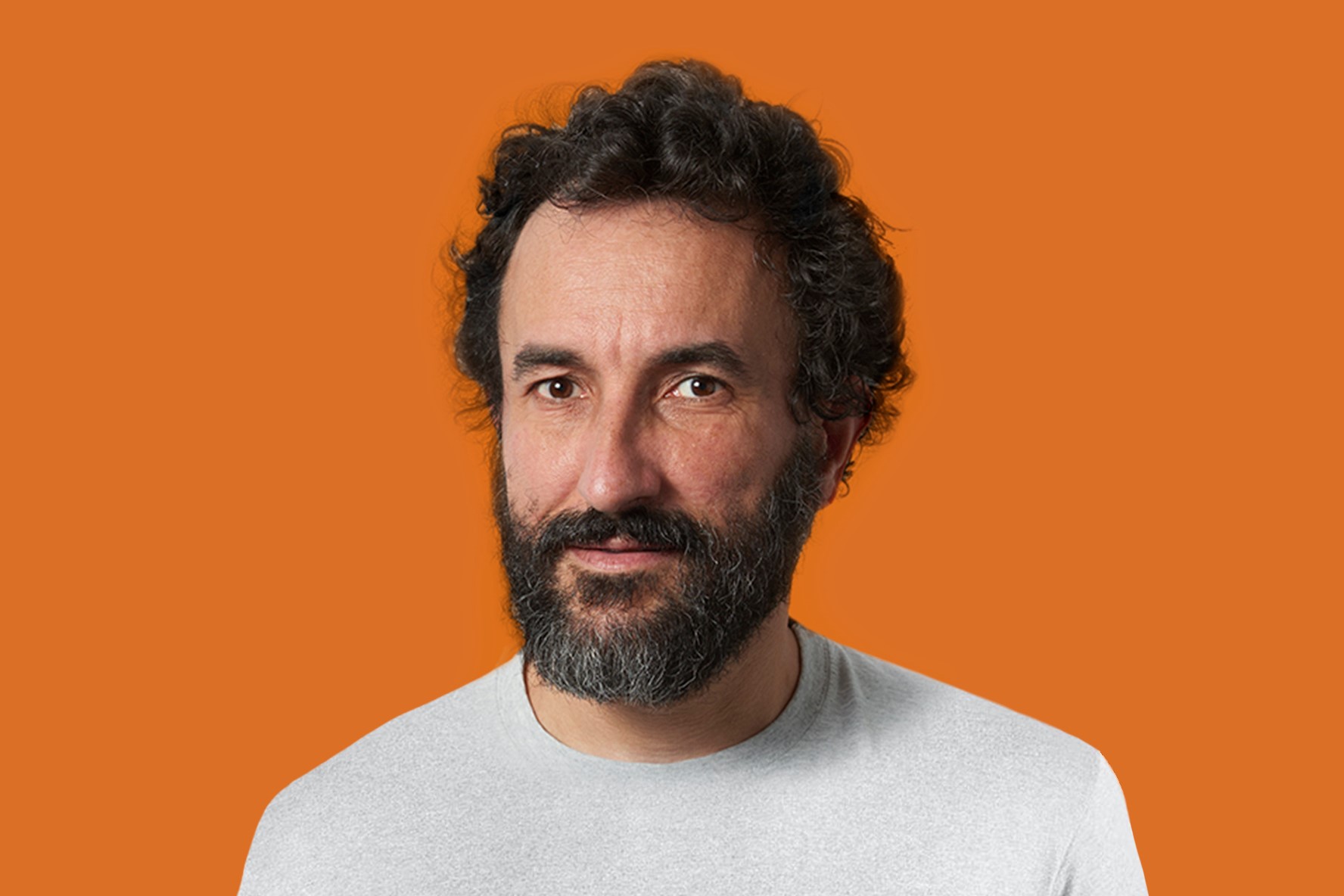}}]{Piergiorgio Cipriano} information systems project manager at Dedagroup, he has been dealing with geomatics and geographic data infrastructures for 25 years. Member of ISO, CEN, INSPIRE working groups for the standardization of geographic data and services, he has worked at CSI-Piemonte, Core Solutions Informatiche, Sinergis and now Dedagroup, with a brief experience of two years at the Joint Research Center of the European Commission. For over 10 years he has been involved in projects financed by European programs (FP7, CIP-PSP, Horizon2020, Horizon Europe, Interreg, CEF-Telecom, EIT Climate KIC) in particular on issues relating to sustainability, energy and mobility in the urban environment.
\end{IEEEbiography}
\vspace{-8 mm}
\begin{IEEEbiography}[{\includegraphics[width=1in,height=1.25in,clip,keepaspectratio]{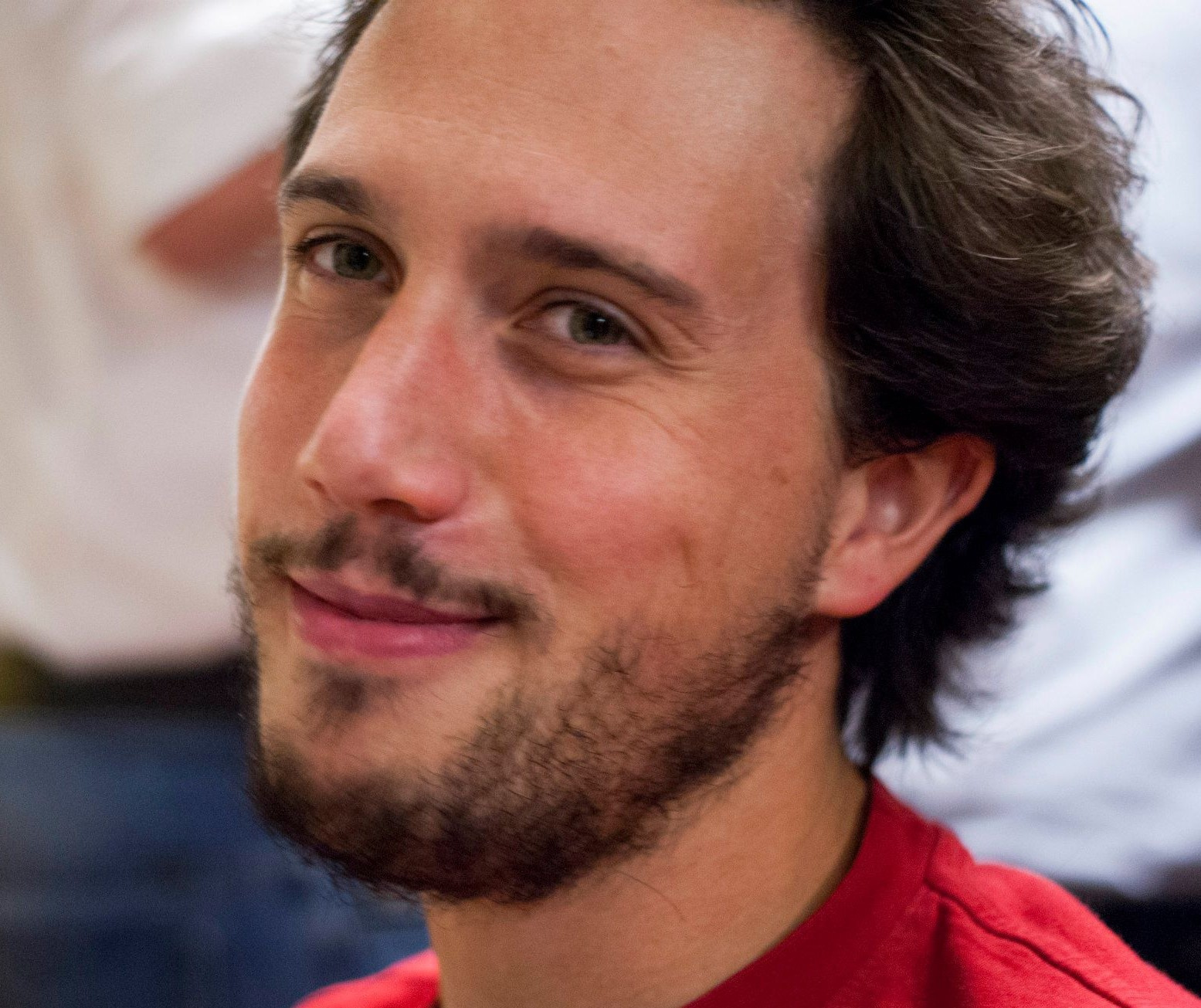}}]{Luca Giovannini} is analyst and technical lead for R\&D projects in the fields of Geomatics and Remote Sensing, with 8 years of experience; Worked on the design of algorithms for vehicle routing and pattern recognition in images; Implemented SAR Interferometry software for ESA Sentinel programme; Currently involved in EU funded projects on the energy and mobility domains of the smart city paradigm, working on the topics of data integration, interoperability and semantics, as well as designing data elaboration procedures.
\end{IEEEbiography}
\vspace{-8 mm}
\begin{IEEEbiography}[{\includegraphics[width=1in,height=1.25in,clip,keepaspectratio]{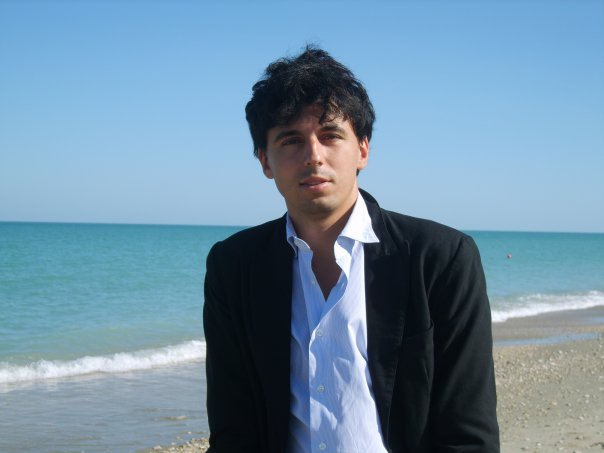}}]{Bruno Lepri} is a senior researcher at Fondazione Bruno Kessler (Trento, Italy) where he leads the Mobile and Social Computing Lab (MobS). He has recently launched the Center for Computational Social Science and Human Dynamics, a joint initiative between Fondazione Bruno Kessler and the University of Trento. Since July 2022, he is the Chief Scientific Officer of Ipazia, a new company active on AI solutions for financial services and energy management. From 2019 to 2022, Bruno was also the Chief AI Scientist of ManpowerGroup where he has collaborated with the global innovation team on AI projects for recruitment and HR management. Bruno is also a senior research affiliate at Data-Pop Alliance, the first think-tank on big data and development co-created by the Harvard Humanitarian Initiative, MIT Media Lab, Overseas Development Institute, and Flowminder. Finally, he has co-founded Profilio, a startup on AI-driven psychometric analysis. In 2010 he won a Marie Curie Cofund postdoc fellowship and he has held a 3 year postdoc position at the MIT Media Lab. He holds a Ph.D. in Computer Science from the University of Trento. His research interests include computational social science, personality computing, network science, and machine learning. His research has received attention from several international press outlets and obtained the 10-year impact award at MUM 2021, the James Chen Annual Award for the best 2016 UMUAI paper, and the best paper award at ACM Ubicomp 2014.
\end{IEEEbiography}
\vspace{-13 cm}
\begin{IEEEbiography}[{\includegraphics[width=1in,height=1.25in,clip,keepaspectratio]{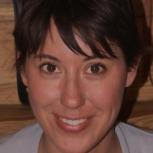}}]{Annapaola Marconi} received her Ph.D. in Computer Science at the University of Trento in 2008. She is currently Senior Researcher at Fondazione Bruno Kessler (FBK), where she directs the Motivational Digital Systems (MoDiS) research unit.  Her research interests include distributed adaptive systems, persuasive and motivational systems based on gamification and gameful design techniques. She is actively involved in the implementation and dissemination of initiatives and projects aimed at applying the research results in concrete applications that can create an impact on the local area and improve the quality of life of its citizens.You can contact the author at \url{marconi@fbk.eu}.
\end{IEEEbiography}

\end{document}